\documentstyle[psfig,graphics]{mn} 
\title [Kinematics of W UMa Stars]
{Kinematics of W UMa-type binaries and evidences on the two types
of formation
}
\author[Bilir et al.]
       {S.~Bilir,$^1 \thanks{E-mail: sbilir@istanbul.edu.tr}$
        Y.~Karata\c{s}$^1$, O.~Demircan$^{2}$\thanks{Visiting Astronomer, 
        Istanbul University Science Faculty, Department of Astronomy and Space
        Sciences} and Z. Eker$^{2}$\footnotemark[2]\\
  $^1$Istanbul University Science Faculty, Department of Astronomy and Space
      Sciences, 34119, University-Istanbul, Turkey\\
  $^2$\c{C}anakkale University, Ulup\i nar Astrophysical Observatory, 
      17100 \c{C}anakkale, Turkey\\}

\pagerange{\pageref{firstpage}--\pageref{lastpage}}
\pubyear{2004}

\begin{document}

\maketitle

\label{firstpage}

\begin{abstract}

The kinematics of 129 W UMa binaries is studied and its implications on the 
contact binary evolution is discussed. The sample is found to be heterogeneous 
in the velocity space that kinematically younger and older contact binaries 
exist in the sample. Kinematically young (0.5 Gyr) sub-sample (MG) is formed by 
selecting the systems which are satisfying the kinematical criteria of moving 
groups. After removing the possible MG members and the systems which are known to 
be members of open clusters, the rest of the sample is called Field Contact 
Binaries (FCB). The FCB has further divided into four groups according to 
the orbital period ranges. Then a correlation has been found in the sense that 
shorter period less massive systems have larger velocity dispersions than the 
longer period more massive systems. Dispersions in the velocity space indicates 
5.47 Gyr kinematical age for the FCB group. Comparing with the field chromospherically 
active binaries (CAB), presumably detached binary progenitors of the contact 
systems, the FCB appears to be 1.61 Gyr older. Assuming an equilibrium 
in the formation and destruction of CAB and W UMa systems in the Galaxy, this age 
difference is treated as empirically deduced lifetime of the contact stage. Since 
the kinematical ages (3.21, 3.51, 7.14 and 8.89 Gyr) of the four sub groups of 
FCB are much longer than the 1.61 Gyr lifetime of the contact stage, the 
pre-contact stages of FCB must dominantly be producing the large dispersions. 
The kinematically young (0.5 Gyr) MG group covers the same total mass, period and 
spectral ranges as the FCB. But, the very young age of this group does not leave 
enough room for pre-contact stages, thus it is most likely that those systems were 
formed in the beginning of the main-sequence or during the pre-main-sequence 
contraction phase, either by a fission process (Roxburgh\ 1966) or most probably by 
fast spiraling in of two components in a common envelope. 

\end{abstract}


\section{Introduction}

The low mass contact binaries, popularly known as W Ursae Majoris stars, are 
easily recognized by their eclipsing light curves with equal (or nearly 
equal) minima as wide as to touch one another. With a small orbit implied 
by a short orbital period less than a day, the binary components fill their 
Roche lobes so that the dumbbell-shaped stars touch each other at the inner 
Langrangian point. The W UMa systems have spectral types ranging from late A to 
middle K and they are located near or just above the main sequence. 
The strong tidal interaction synchronized their rotation thus the period of 
rotation is equal to the orbital revolution. The rapid rotation and 
convective atmospheres are primary causes to their observed chromospheric and 
coronal emissions as well as to the presence of large starspots which are the 
manifestations of strong, dynamo-generated magnetic activity. 

The rapid rotation and quickly changing radial velocity, on the other hand, 
is a serious obstacle to obtain a reliable radial velocity from the highly 
broadened and blended spectral lines of numerous W UMa systems. Therefore, 
forming a complete sample for studying galactic space motions and kinematics of 
W UMa stars is limited with the availability of the radial velocity measurements.
From the identified 374 W UMa binaries (Kukarkin et al.\ 1971; 
van't Veer\ 1975) in the 3rd General Catalog of Variable Stars (GCVS), Guinan \& 
Bradstreet\ (1988) were able to collect only 34 systems with a measured center 
of mass velocity. The number of stars classified as EW (eclipsing binaries of 
W UMa type) increased to 561 in the 4th edition of GCVS. But there were only 
78 systems available with complete physical parameters which were studied 
statistically by Maceroni \& van't Veer\ (1996). Only 37 out 78 had 
spectroscopic mass ratio determined from the radial velocities. Finally, 
Aslan et al.\ (1999) counted 125 W UMa binaries in the Hipparcos Catalogue, but 
there were only 34 binaries from which the space motions can be computed. The revised 
4th edition of GCVS now contains 751 W UMa. Thanks to the dedication of David Dunlap 
Observatory (DDO), where 54 per cent of current radial velocities came from, and 
Dominion Astrophysical Observatory (DAO) with 20 per cent contribution, and the 
improvements of spectroscopic techniques for measuring radial velocities that a total 
of 129 W UMa systems were collected for this study. Nearly four times the number and 
the improved accuracy of data justified us to repeat previous attempts of studying 
kinematics of W UMa binaries.

With a small sample of less accurate data, Guinan \& Bradstreet\ (1988) estimated a 
kinematical age of 8-10 Gyr for W UMa systems. If this age is compared to the 
$\sim5$ Gyr average age of chromospherically active binaries (CAB), 
classical RS CVn and BY Dra systems (Eker\ 1992), W UMa stars appear two times older. 
Nevertheless, this age of W UMa binaries was found consistent with the theory of contact 
binary formation from the detached binary progenitors, presumably from the short period 
($P<5$ days) RS CVn binaries, by a mechanism involving angular momentum loss and 
orbital period decrease previously suggested by Huang\ (1967), Mestel\ 
(1968), Okamoto \& Sato\ (1970), van't Veer\ (1979) and Vilhu \& Raunen\ (1980). 
But, according to the space velocities and dispersions computed by Aslan et al.\
(1999), W UMa binaries cannot be older than RS CVn systems. This means that 
W UMa binaries could not have evolved from detached binaries unless the time to 
reach contact stage is very short. 

With no pre-contact stage, there is an opposing theory which predicts the formation 
of contact binaries directly by a fission process (Roxburgh\ 1966) at the end of the 
pre-main sequence contraction. Considering the brief lifetime of the contact stage 
$(0.1<t_{contact}<1$ Gyr) estimated by Guinan \& Bradstreet\ (1988), this theory would 
fail to produce older populations of W UMa binaries unless the angular momentum of 
contact binaries is conserved. However, with the conservation of the orbital angular 
momentum, this theory would refuse the formation of the contact binaries from the 
detached binary progenitors.

Apparently, an empirical evidence implying whether the total orbital 
angular momentum is conserved or not within the evolutionary time scale is 
extremely important to reveal the actual working model. Inspired by the 
predictions of Demircan\ (1999), Karata\c{s} et al.\ (2004) presented such an evidence 
recently from the kinematics of CAB sample containing RS CVn and 
BY Dra stars. The evidence of orbital period decrease, connected to mass and 
angular momentum loss, was shown by comparing the total mass ($M_{1}+M_{2}$) and 
period histograms between the kinematically younger (age 0.95 Gyr) and the older 
(age 3.86 Gyr) CAB sub samples. The orbital period decrease was also indicated by 
the fact that the shorter orbital period systems were found older than the longer 
orbital period systems. According to a preliminary investigation of Demircan et al.\ 
(2004), the period decrease is exponential in time so that the active detached binaries,   
up to about 5 days orbital period, could change into contact form within less than about 
5 Gyr. Therefore, the formation of W UMa stars from detached binaries is evidently 
occurring. Nevertheless, this does not mean this is the only mechanism to form contact 
binaries. It may still possible that a limited fraction W UMa systems, possibly the very 
young ones, may be forming by the fission process.    

In the present work, the kinematics of W UMa binaries is studied and its 
implications on the contact binary evolution is discussed. The present sample has 
been found heterogeneous like the CAB sample in a sense that there are younger 
($\sim1$ Gyr age) and older ($\sim6$ Gyr age) W UMa binaries covering the same range 
of spectral types and orbital periods. The evidence of mass and angular momentum 
loss causing orbital period decrease during contact binary evolution has been 
investigated for a preliminary step as done by CAB sample by 
Karata\c{s} et al.\ (2004) by comparing the total mass ($M_{1}+M_{2}$) and period 
histograms between the younger and the older sub samples.

\section{Data}
Among the 751 recorded W UMa binaries in the recently revised 4th edition of GCVS, 
129 systems were found to have radial velocities. After compiling 
the rest of the basic data (parallaxes and proper motions), the sample is 
listed in Table 1, which is self explanatory, with the columns: order number, name,
HD and Hipparcos reference numbers, celestial coordinates (ICRS 2000), proper 
motions, parallaxes and radial velocities. The associated standard errors are given besides 
the related parameter. The reference numbers in the last column are 
separated into three fields with semicolons to indicate from where the basic data 
were taken. The two or more reference numbers in a field indicate the sub fields, 
that is, there are multiple references. 

\begin{table*}
\caption{Hipparcos astrometric and radial velocity data of the W UMa.}
{\scriptsize
\begin{tabular}{rrrrrrrrrrrrrrr}
\hline
ID &   Name &  HD &  HIP & $\alpha$ (2000) & $\delta$ (2000)& \multicolumn{2} {c} {$\mu_{\alpha}cos\delta$} & \multicolumn{2} {c} {$\mu_{\delta}$} &
\multicolumn{2} {c} {$\pi$} &  \multicolumn{2} {c} {$\gamma$} &  References\\ 
   &        &     &      & ($h~~~m~~~s$)        & ($\degr~~~\arcmin~~~''$)       &  \multicolumn{2} {c} {(mas yr$^{-1}$)}& \multicolumn{2} {c} {(mas yr$^{-1}$)} & \multicolumn{2} {c} {(mas)} &  \multicolumn{2} {c} {(km s$^{-1}$)} &\\   
\hline
  1 &     AQ Tuc &       1372 &       1387 & 00 17 21.51 &$-$71 54 56.84& 28.37 & 1.20 & - 9.37& 1.10 &  2.84&  0.36 &  20.60&  0.27 &    (1;5;6) \\
  2 &     BH Cas &            &            & 00 21 40.30 & +59 12 06.60 & -7.43 & 2.50 & -32.18& 2.50 &  3.37&  0.42 & -23.70&  2.40 &    (2;5;7) \\
  3 &     DZ Psc &            &            & 00 36 27.93 & +21 32 13.60 & -6.30 & 2.50 &  14.90& 2.50 &  4.83&  0.60 & -21.66&  2.15 &    (2;5;8) \\
  4 &   V523 Cas &            &            & 00 40 06.26 & +50 14 15.53 &146.00 & 2.60 & -71.10& 2.50 & 12.44&  1.56 &  -2.54&  0.90 &(2;5;8,9,10)\\
  5 &     AQ Psc &       8152 &       6307 & 01 21 03.56 & +07 36 21.63 & -8.72 & 1.38 &  14.58& 0.73 &  8.03&  1.29 & -12.90&  0.40 &   (1;1;11) \\
  6 &     AE Phe &       9528 &       7183 & 01 32 32.93 &$-$49 31 41.29&150.93 & 0.80 & -54.47& 0.65 & 20.49&  0.81 &  -9.50&  2.50 &   (1;1;12) \\
  7 &     TW Cet &            &       8447 & 01 48 54.14 &$-$20 53 34.60& 44.82 & 3.69 & -15.91& 2.18 &  9.91&  3.50 &  20.00&  25.0 &   (1;1;13) \\
  8 &   V776 Cas &            &       8821 & 01 53 23.44 & +70 02 33.44 & 19.44 & 1.51 &   4.32& 1.07 &  4.86&  1.56 & -24.71&  0.69 &   (1;1;14) \\
  9 &     QX And &            &            & 01 57 57.10 & +37 48 23.00 & 12.10 & 2.50 & -11.50& 2.50 &  2.63&  0.23 &  11.70&  0.27 &  (2;15;15) \\
 10 &     SS Ari &            &            & 02 04 18.41 & +24 00 02.23 & 25.60 & 2.50 & -10.70& 2.50 &  4.71&  0.59 &  -3.20&  1.30 &   (2;5;16) \\
 11 &     GZ And &            &      10270 & 02 12 14.07 & +44 39 36.28 & -15.40& 2.50 &  -0.60& 2.50 &  6.15&  0.77 &   0.00&  1.20 &   (2;5;11) \\
 12 &   V376 And &      15922 &      12039 & 02 35 11.63 & +49 51 37.20 & 49.40 & 0.86 &  -8.28& 0.66 &  5.07&  0.89 &  22.83&  0.89 &   (1;1;14) \\
 13 &     EE Cet &      17613 &      13199 & 02 49 52.38 & +08 56 22.63 & 45.90 & 5.00 &  -1.00& 4.90 &  7.29&  0.91 &   1.60&  0.93 &   (3;5;17) \\
 14 &     UX Eri &            &      14699 & 03 09 52.74 & -06 53 33.56 & 24.13 & 2.85 &  -3.80& 1.91 &  6.57&  2.84 &  12.79&  1.09 &   (1;1;18) \\
 15 &     EQ Tau &            &            & 03 48 13.43 & +22 18 50.94 & 68.40 & 2.50 & -29.60& 2.50 &  6.97&  0.87 &  71.95&  1.22 &   (2;5;14) \\
 16 &     BL Eri &            &            & 04 11 48.18 &$-$11 47 26.52& 11.70 & 2.70 &   1.00& 2.70 &  2.77&  0.35 &  40.00&  1.51 &   (3;5;19) \\
 17 &     YY Eri &      26609 &      19610 & 04 12 08.85 &$-$10 28 09.96&-111.87& 1.46 &-114.52& 0.96 & 17.96&  1.20 & -15.00&  1.50 &   (1;1;20) \\
 18 &     AO Cam &            &            & 04 28 13.64 & +53 02 44.52 & 14.80 & 2.50 & -23.30& 2.50 &  9.20&  1.15 & -13.17&  0.83 &   (2;5;18) \\
 19 &     RZ Tau &     285892 &      21467 & 04 36 37.66 & +18 45 17.80 & -26.31& 1.76 &  -6.42& 1.28 &  5.74&  1.85 &   5.00&  50.0 &   (1;1;13) \\
 20 &     DN Cam &      29213 &      21913 & 04 42 46.24 & +72 58 41.87 & -17.90& 0.80 & -12.98& 0.50 &  4.49&  0.89 &   6.04&  0.98 &   (1;1;14) \\
 21 &   V410 Aur &     280332 &      23337 & 05 01 10.84 & +34 30 26.57 & 9.04  & 4.97 &  -5.33& 2.99 &  4.77&  5.39 &  36.94&  3.10 &    (1;1;8) \\
 22 &   V402 Aur &     282719 &      23433 & 05 02 14.74 & +31 15 49.29 & 9.68  & 1.58 &  -5.84& 1.05 &  7.01&  1.31 &  40.82&  0.93 &   (1;1;21) \\
 23 &     AP Dor &      33474 &      23793 & 05 06 45.09 &$-$59 03 03.45& 56.68 & 1.23 &  77.85& 0.73 &  4.66&  0.92 &   6.80&  0.97 &   (1;1;22) \\
 24 &  V1363 Ori &     289949 &      23809 & 05 07 02.11 &$-$00 47 32.64& 21.99 & 2.27 & -42.58& 1.50 &  9.47&  2.36 &  37.89&  2.02 &   (1;1;21) \\
 25 &     ER Ori &            &      24156 & 05 11 14.50 &$-$08 33 24.66& 7.20  & 1.71 & -25.73& 1.14 &  6.24&  0.78 &  37.90&  3.30 &   (1;5;23) \\
 26 &     RW Dor &     269320 &      24763 & 05 18 32.54 &$-$68 13 32.76& 96.63 & 2.57 & 55.06 & 1.86 &  8.91&  1.97 &  66.50&  1.45 &   (1;1;24) \\
 27 &   V781 Tau &     248087 &      27562 & 05 50 13.12 & +26 57 43.37 & -81.59& 1.51 & -85.08& 0.77 & 12.31&  1.35 &  24.40&  1.30 &   (1;1;25) \\
 28 &     AH Aur &     256902 &      30618 & 06 26 04.90 & +27 59 56.40 & 15.21 & 2.23 & -12.72& 1.36 &  6.18&  2.05 &  31.95&  1.45 &   (1;1;26) \\
 29 &     QW Gem &     264672 &      32845 & 06 50 46.07 & +29 27 11.34 & 16.57 & 8.12 & -22.58& 2.97 &  4.09&  5.42 &  -0.90&  1.39 &    (1;1;8) \\
 30 &   V753 Mon &      54975 &      34684 & 07 10 57.85 &$-$03 52 43.19& -13.56& 0.98 &  -9.14& 0.66 &  5.23&  1.04 &  38.58&  0.83 &   (1;1;18) \\
 31 &     TY Pup &      60265 &      36683 & 07 32 46.23 &$-$20 47 31.07& -24.05& 1.09 &  42.09& 0.76 &  4.67&  0.58 &  28.00&  30.0 &   (1;5;27) \\
 32 &     FG Hya &            &      41437 & 08 27 03.94 & +03 30 52.33 & 6.88  & 1.92 & -63.68& 1.40 &  2.92&  1.90 &  -5.70&  1.20 &   (1;1;11) \\
 33 &     TX Cnc &            &            & 08 40 01.67 & +18 59 58.72 & -48.20& 4.00 & -19.00& 3.90 &  5.21&  0.79 &  29.00&  2.12 &  (3;68;28) \\
 34 &     AH Cnc &            &            & 08 51 37.87 & +11 50 57.20 & -7.80 & 2.50 &  -9.10& 2.50 &  1.20&  0.10 &  15.00&  1.89 &  (2;67;29) \\
 35 &     UV Lyn &            &      44455 & 09 03 24.12 & +38 05 54.60 & -71.67& 1.79 &  18.85& 1.02 &  8.16&  1.62 &  -0.30&  1.30 &   (1;1;11) \\
 36 &     FN Cam &      79886 &      46005 & 09 22 58.04 & +77 13 10.95 & -4.31 & 1.13 & -28.56& 0.74 &  3.62&  0.95 &  12.96&  1.11 &   (1;1;14) \\
 37 &     EZ Hya &            &            & 09 26 41.06 &$-$13 45 06.41& -30.80& 3.30 &  -0.30& 2.50 &  5.43&  0.68 &  16.00&  0.44 &   (4;5;30) \\
 38 &      S Ant &      82610 &      46810 & 09 32 18.39 &$-$28 37 39.97& -88.84& 0.55 &  43.79& 0.49 & 13.30&  0.71 &  -5.00&  1.80 &   (1;1;31) \\
 39 &      W UMa &      83950 &      47727 & 09 43 45.47 & +55 57 09.08 & 15.55 & 1.03 & -27.35& 0.62 & 20.17&  1.05 & -29.60&  1.22 &   (1;1;32) \\
 40 &     AA UMa &            &            & 09 46 59.29 & +45 45 56.40 & -1.00 & 2.50 &  -4.80& 2.50 &  3.19&  0.40 & -34.80&  1.90 &   (2;5;33) \\
 41 &     RT Lmi &            &            & 09 49 48.32 & +34 27 15.40 & 3.60  & 2.50 &  -6.00& 2.50 &  3.07&  0.38 & -10.30&  1.74 &   (2;5;18) \\
 42 &     XY Leo &            &      49136 & 10 01 40.43 & +17 24 32.70 & 56.70 & 1.68 & -58.27& 0.71 & 15.86&  1.80 & -50.00&  2.00 &   (1;1;34) \\
 43 &     XZ Leo &            &      49204 & 10 02 34.19 & +17 02 47.14 & -16.10& 1.78 &   1.85& 0.81 &  2.59&  0.32 &  -2.03&  1.43 &   (1;5;26) \\
 44 &      Y Sex &      87079 &      49217 & 10 02 47.96 & +01 05 40.35 & 0.67  & 3.84 &  -2.67& 2.05 &  9.02&  3.38 &   9.80&  2.00 &   (1;1;28) \\
 45 &     ET Leo &      91386 &      51677 & 10 33 25.79 & +17 34 27.43 & -37.71& 1.53 &  -7.21& 0.80 & 13.90&  1.44 &  21.46&  0.78 &   (1;1;17) \\
 46 &     UZ Leo &            &      52249 & 10 40 33.19 & +13 34 00.86 & -21.60& 1.49 &  -1.10& 0.89 &  6.27&  1.59 &  -8.18&  1.12 &   (1;1;26) \\
 47 &     EX Leo &      93077 &      52580 & 10 45 06.77 & +16 20 15.68 & -26.51& 1.23 & -37.11& 0.75 &  9.84&  1.11 & -11.05&  1.10 &   (1;1;35) \\
 48 &     VY Sex &      93917 &            & 10 50 29.72 &$-$02 41 43.05& -21.40& 2.90 & -88.60& 2.20 &  7.34&  0.92 &  17.47&  0.79 &    (4;5;8) \\
 49 &     AM Leo &            &      53937 & 11 02 10.89 & +09 53 42.69 & -9.98 & 4.54 & -29.53& 3.26 & 13.03&  3.64 &   5.60&  1.20 &   (1;1;36) \\
 50 &     VW Lmi &      95660 &      54003 & 11 02 51.91 & +30 24 54.71 & 12.21 & 0.98 &  -3.92& 0.66 &  8.04&  0.90 &   5.00&  0.75 &   (1;1;37) \\
 51 &     AP Leo &            &      54188 & 11 05 05.02 & +05 09 06.42 & 89.94 & 2.05 & -59.58& 1.17 &  8.26&  1.70 & -25.10&  2.10 &   (1;1;11) \\
 52 &     HN UMa &            &      55030 & 11 15 56.38 & +37 38 35.30 & 56.12 & 1.28 & -35.67& 0.90 &  5.81&  1.39 & -37.11&  0.63 &    (1;1;8) \\
 53 &     AW UMa &      99946 &      56109 & 11 30 04.32 & +29 57 52.67 & -82.35& 0.87 &-199.74& 0.68 & 15.13&  0.90 & -17.00&  1.57 &   (1;1;38) \\
 54 &     TV Mus &     310730 &            & 11 39 57.76 &$-$64 48 59.29& 27.00 & 5.20 &  -6.80& 3.20 &  4.32&  0.54 &   3.10&  1.00 &   (4;5;39) \\
 55 &   V752 Cen &     101799 &      57129 & 11 42 48.08 &$-$35 48 57.51& -53.65& 1.63 & -24.28& 0.93 &  9.51&  1.47 &  29.10&  0.82 &   (1;1;40) \\
 56 &     AG Vir &     104350 &      58605 & 12 01 03.50 & +13 00 30.02 & -2.46 & 1.46 & -17.64& 0.56 &  5.02&  0.63 &  -5.91&  0.73 & (1;5;41,58)\\
 57 &     HX UMa &     104425 &      58648 & 12 01 33.15 & +43 02 29.94 & 65.86 & 2.56 & -21.34& 2.15 &  6.68&  3.01 & -19.88&  1.11 &    (1;1;8) \\
 58 &     CC Com &            &            & 12 12 06.20 & +22 31 58.00 &-120.00& 2.50 &  19.30& 2.50 & 12.49&  1.56 & -10.20&  1.31 &   (2;5;42) \\
 59 &     AH Vir &     106400 &      59683 & 12 14 21.00 & +11 49 09.40 & 47.74 & 3.03 &-107.26& 1.39 & 10.86&  3.11 &   6.60&  0.90 &   (1;1;43) \\
 60 &     II UMa &     109247 &      61327 & 12 32 54.84 & +54 47 42.89 & -42.50& 1.57 & -12.12& 1.45 &  5.04&  1.81 &  -8.02&  1.10 &   (1;1;17) \\
 61 &     RW Com &            &      61243 & 12 33 00.28 & +26 42 58.38 &-125.45& 3.46 & -36.05& 1.89 & 11.45&  2.45 & -53.00&  1.15 &   (1;1;44) \\
 62 &     RZ Com &            &      61414 & 12 35 05.06 & +23 20 14.02 & 17.44 & 1.79 & -11.01& 1.18 &  4.70&  0.59 &  -1.80&  1.67 &   (1;5;28) \\
 63 &     SX Crv &     110139 &      61825 & 12 40 15.04 &$-$18 48 00.93& 37.93 & 1.21 &  -5.63& 0.92 & 10.90&  1.21 &   8.71&  0.94 &   (1;1;14) \\
 64 &     BI CVn &            &      63701 & 13 03 16.41 & +36 37 00.59 & -6.38 & 1.23 & -16.68& 1.04 &  5.31&  1.65 &  -5.30&  1.80 &   (1;1;45) \\
 65 &     KZ Vir &     114726 &      64433 & 13 12 22.87 & +02 39 13.78 & -8.53 & 0.98 &   0.94& 0.48 &  6.53&  0.93 &  -4.50&  0.48 &   (1;1;14) \\
 \hline
\end{tabular}  
}
\end{table*}

\begin{table*}
\contcaption{}
{\scriptsize
\begin{tabular}{rrrrrrrrrrrrrrr}
\hline
ID &   Name &  HD &  HIP & $\alpha$ (2000) & $\delta$ (2000)& \multicolumn{2} {c} {$\mu_{\alpha}cos\delta$} & \multicolumn{2} {c} {$\mu_{\delta}$} &
\multicolumn{2} {c} {$\pi$} &  \multicolumn{2} {c} {$\gamma$} &  References\\ 
   &        &     &      & ($h~~~m~~~s$)        & ($\degr~~~\arcmin~~~''$)       &  \multicolumn{2} {c} {(mas yr$^{-1}$)}& \multicolumn{2} {c} {(mas yr$^{-1}$)} & \multicolumn{2} {c} {(mas)} &  \multicolumn{2} {c} {(km s$^{-1}$)} &\\   
\hline
 66 &     KR Com &     115955 &      65069 & 13 20 15.78 & +17 45 56.99 & -6.70 & 0.78 & -82.30& 0.47 & 13.07&  0.87 &  -7.86&  0.38 &   (1;1;17) \\
 67 &     HT Vir &     119931 &      67186 & 13 46 06.75 & +05 06 56.27 &-107.79& 3.24 & -11.16& 1.68 & 15.39&  2.72 & -23.38&  0.68 &   (1;1;35) \\
 68 &     XY Boo &            &      67431 & 13 49 11.57 & +20 11 24.59 & -23.32& 1.87 &  26.91& 0.89 &  2.94&  1.70 &   2.11&  2.02 &   (1;1;28) \\
 69 &   V757 Cen &     120734 &      67682 & 13 51 55.74 &$-$36 37 24.82&-121.51& 1.20 & -61.89& 0.94 & 14.18&  1.10 &  50.00&  2.50 &   (1;1;46) \\
 70 &     RR Cen &     124689 &      69779 & 14 16 57.22 &$-$57 51 15.65& -52.69& 0.77 & -22.37& 0.57 &  9.76&  0.85 & -16.00&  0.41 &   (1;1;30) \\
 71 &     VW Boo &            &      69826 & 14 17 26.03 & +12 34 03.45 & 5.45  & 2.09 & -48.79& 0.95 &  3.44&  1.90 &  23.90&  0.82 &   (1;1;47) \\
 72 &     NN Vir &     125488 &      70020 & 14 19 37.74 & +05 53 46.66 & -54.17& 1.09 & -19.63& 0.60 &  9.48&  1.14 &  -6.24&  0.65 &   (1;1;26) \\
 73 &     EF Boo &     234150 &      71107 & 14 32 30.54 & +50 49 40.70 & -33.23& 1.17 & -37.34& 0.93 &  6.00&  1.06 & -13.16&  1.07 &   (1;1;14) \\
 74 &     CK Boo &     128141 &      71319 & 14 35 03.76 & +09 06 49.39 & 70.91 & 1.18 & -86.58& 0.80 &  6.38&  1.34 &  37.04&  0.75 &   (1;1;26) \\
 75 &     GR Vir &     129903 &      72138 & 14 45 20.26 &$-$06 44 04.13& -84.94& 1.47 &  47.34& 0.99 & 18.83&  1.18 & -71.72&  0.89 &   (1;1;26) \\
 76 &     AC Boo &            &      73103 & 14 56 28.34 & +46 21 44.14 & -9.11 & 1.30 &   8.99& 1.26 &  7.58&  1.27 &  -8.60&  2.20 &   (1;1;36) \\
 77 &     TY Boo &            &            & 15 00 46.94 & +35 07 54.89 & -58.80& 2.50 &  32.30& 2.50 &  5.82&  0.73 & -41.90&  1.55 &   (2;5;48) \\
 78 &     44 Boo &     133640 &      73695 & 15 03 47.30 & +47 39 14.62 &-436.26& 1.36 &  18.93& 0.98 & 78.39&  1.03 & -17.89&  0.40 &   (1;1;35) \\
 79 &     TZ Boo &            &      74061 & 15 08 09.13 & +39 58 12.86 & -89.03& 1.53 &  58.80& 1.42 &  6.76&  1.49 & -36.70&  1.51 &   (1;1;28) \\
 80 &     BW Dra &            &      74368 & 15 11 50.11 & +61 51 41.26 &-153.96& 8.04 &  98.81& 6.44 & 15.32&  5.18 & -60.80&  1.15 &   (1;1;49) \\
 81 &     BV Dra &     135421 &      74370 & 15 11 50.36 & +61 51 25.32 &-162.57& 2.98 &  99.13& 2.47 & 14.86&  2.56 & -61.45&  1.45 &   (1;1;49) \\
 82 &     FI Boo &     234224 &      75203 & 15 22 05.97 & +51 10 55.33 & -36.87& 2.46 &  47.43& 2.29 &  9.52&  2.10 & -30.55&  0.72 &   (1;1;35) \\
 83 &     OU Ser &     136924 &      75269 & 15 22 43.47 & +16 15 40.74 &-387.51& 0.91 &   2.80& 0.77 & 17.31&  0.95 & -64.08&  0.41 &   (1;1;18) \\
 84 &     VZ Lib &            &      76050 & 15 31 51.76 &$-$15 41 00.19& -15.61& 2.55 &  -1.94& 1.75 &  4.92&  1.96 & -31.11&  2.30 &   (1;1;35) \\
 85 &     FU Dra &            &      76272 & 15 34 45.21 & +62 16 44.28 &-255.85& 1.18 &  15.61& 1.11 &  6.25&  1.09 & -11.38&  1.14 &   (1;1;18) \\
 86 &     YY CrB &     141990 &      77598 & 15 50 32.43 & +37 50 07.56 & -74.09& 0.92 &  10.93& 0.71 & 11.36&  0.85 &  -4.58&  1.02 &   (1;1;18) \\
 87 &     AU Ser &            &            & 15 56 49.47 & +22 16 01.59 & -31.40& 2.50 & -48.60& 2.50 &  5.46&  0.68 & -62.90&  1.50 &   (2;5;36) \\
 88 &   V842 Her &            &            & 16 06 02.34 & +50 11 12.10 & -27.90& 2.50 &  20.70& 2.50 &  6.08&  0.76 & -57.98&  1.55 &   (2;5;26) \\
 89 &     BX Dra &            &      78891 & 16 06 17.37 & +62 45 46.10 & -7.80 & 1.11 &   7.55& 0.92 &  3.38&  1.15 & -26.11&  3.43 &   (1;1;21) \\
 90 &   V899 Her &     149684 &      81191 & 16 35 01.96 & +33 12 47.77 & 39.09 & 0.72 & -39.57& 0.55 &  8.06&  0.77 & -16.84&  1.30 &   (1;1;35) \\
 91 &   V502 Oph &     150484 &      81703 & 16 41 20.86 & +00 30 27.37 & -27.95& 1.76 &  20.92& 0.86 & 11.84&  1.17 & -42.56&  0.85 &   (1;1;21) \\
 92 &   V918 Her &     151701 &      82253 & 16 48 24.18 & +17 07 59.63 & -15.88& 0.88 &  14.27& 0.62 &  8.70&  0.93 & -25.72&  0.74 &   (1;1;21) \\
 93 &   V921 Her &     152172 &      82344 & 16 49 31.24 & +47 06 28.81 & 7.46  & 0.98 &   5.15& 0.91 &  2.39&  0.97 & -79.04&  1.05 &    (1;1;8) \\
 94 &   V829 Her &            &            & 16 55 47.88 & +35 10 57.60 & -11.80& 3.10 & -15.60& 2.60 & 13.53&  0.54 & -13.40&  1.00 &  (4;69;11) \\
 95 &  V2357 Oph &            &      82967 & 16 57 16.76 & +10 59 51.38 & 0.51  & 1.80 &-103.60& 1.11 &  5.37&  2.01 & -19.12&  1.64 &    (1;1;8) \\
 96 &     AK Her &     155937 &      84293 & 17 13 57.82 & +16 21 00.61 & 26.24 & 2.23 & -47.63& 1.99 & 10.47&  2.77 & -13.00&  15.0 &   (1;1;50) \\
 97 &   V728 Her &            &            & 17 18 04.46 & +41 50 40.40 & -14.40& 2.50 & -10.10& 2.50 &  2.41&  0.30 &  34.40&  0.80 &   (2;5;51) \\
 98 &     GM Dra &     238677 &      84837 & 17 20 21.88 & +57 58 27.00 & 77.21 & 1.13 & -59.05& 1.02 & 10.16&  0.88 &   9.12&  1.63 &   (1;1;17) \\
 99 &  V2377 Oph &     159356 &      85944 & 17 33 56.05 & +08 09 57.81 & -12.87& 1.00 & -59.16& 0.72 & 10.09&  1.22 & -25.79&  0.38 &   (1;1;35) \\
100 &   V535 Ara &     159441 &      86306 & 17 38 05.55 &$-$56 49 17.29& -76.64& 0.90 & -49.51& 0.50 &  8.87&  0.90 & -17.60&  0.60 &   (1;1;52) \\
101 &  V2388 Oph &     163151 &      87655 & 17 54 14.17 & +11 07 50.01 & -70.81& 0.87 &-163.11& 0.50 & 14.72&  0.81 & -25.88&  0.52 &   (1;1;17) \\
102 &   V566 Oph &     163611 &      87860 & 17 56 52.41 & +04 59 15.32 & 54.04 & 1.17 &  73.92& 0.77 & 13.98&  1.11 & -42.95&  1.13 & (1;1;53,66)\\
103 &   V972 Her &     164078 &      87958 & 17 58 05.00 & +32 38 53.02 & -34.42& 0.51 & -28.43& 0.46 & 16.25&  0.61 &   3.98&  0.46 &   (1;1;17) \\
104 &   V508 Oph &            &      88028 & 17 58 48.62 & +13 29 46.27 & -14.62& 1.88 & -45.07& 1.38 &  7.68&  2.14 & -36.40&  2.05 &   (1;1;54) \\
105 &     EF Dra &            &            & 18 05 30.48 & +69 45 15.71 & -13.60& 2.70 & -24.40& 2.40 &  5.74&  0.72 & -42.20&  3.50 &   (4;5;11) \\
106 &   V839 Oph &     166231 &      88946 & 18 09 21.27 & +09 09 03.63 & -32.48& 1.33 &   3.23& 1.03 &  8.09&  1.44 & -63.76&  0.32 &   (1;1;55) \\
107 &    eps Cra &     175813 &      93174 & 18 58 43.38 &$-$37 06 26.48&-132.25& 1.46 &-110.45& 0.55 & 33.43&  0.92 &  57.90&  1.20 &   (1;1;56) \\
108 &   V401 Cyg &            &      95816 & 19 29 20.27 & +30 24 28.48 & -4.96 & 8.70 & -12.59& 5.99 & 11.55&  8.04 &  25.53&  2.14 &   (1;1;17) \\
109 &  V2082 Cyg &     183752 &      95833 & 19 29 30.07 & +36 17 14.93 & 20.17 & 0.53 &  94.83& 0.50 & 11.04&  0.56 & -34.12&  0.58 &   (1;1;21) \\
110 &   V417 Aql &            &      96349 & 19 35 24.12 & +05 50 17.66 & -1.90 & 2.86 & -24.45& 1.50 &  7.65&  2.35 & -14.20&  1.40 &   (1;1;11) \\
111 &     OO Aql &     187183 &            & 19 48 12.65 & +09 18 32.38 & 65.70 & 1.60 &  -7.90& 1.60 &  6.52&  0.82 & -45.20&  0.90 &   (4;5;57) \\
112 &     VW Cep &     197433 &     101750 & 20 37 21.54 & +75 36 01.46 & 309.23& 1.12 & 539.47& 0.97 & 36.16&  0.97 &  -7.90&  1.00 &   (1;1;59) \\
113 &     LS Del &     199497 &            & 20 57 10.29 & +19 38 59.23 & 145.40& 1.90 &  83.40& 1.80 & 12.71&  1.59 & -25.90&  1.40 &   (4;5;11) \\
114 &  V2150 Cyg &     202924 &     105162 & 21 18 10.88 & +30 35 21.58 & 2.58  & 1.27 &  -6.25& 0.91 &  4.73&  1.59 & -12.82&  0.45 &   (1;1;35) \\
115 &     HV Aqr &            &            & 21 21 24.81 &$-$03 09 36.88& -63.50& 1.90 & -92.30& 1.80 &  6.99&  0.87 &  -3.77&  1.25 &   (4;5;18) \\
116 &  V1073 Cyg &     204038 &     105739 & 21 25 00.36 & +33 41 14.94 & -4.70 & 2.50 & -21.00& 2.50 &  5.44&  0.95 &  -0.80&  1.10 &   (2;1;60) \\
117 &     KP Peg &     204215 &     105882 & 21 26 41.13 & +13 41 17.75 & -2.31 & 2.40 & -10.65& 1.20 &  4.37&  1.67 &   5.52&  1.03 &   (1;1;21) \\
118 &     DK Cyg &            &     106574 & 21 35 02.67 & +34 35 45.40 & -28.37& 1.24 & -19.29& 0.93 &  4.42&  1.78 &  -5.53&  1.48 &   (1;1;26) \\
119 &     BX Peg &            &            & 21 38 49.39 & +26 41 34.20 & 7.10  & 2.50 & -11.90& 2.50 &  6.13&  0.77 &  25.00&  3.00 &   (2;5;61) \\
120 &   V445 Cep &     210431 &     109191 & 22 07 10.89 & +72 22 22.26 & 83.91 & 0.51 &  43.90& 0.42 &  8.95&  0.50 &  40.69&  0.95 &   (1;1;21) \\
121 &     BB Peg &            &     110493 & 22 22 56.89 & +16 19 27.84 & 19.28 & 2.66 & -24.66& 1.85 &  3.02&  2.26 & -28.90&  1.40 &   (1;1;11) \\
122 &   V335 Peg &     216417 &     112960 & 22 52 36.18 & +10 13 13.19 & 166.57& 0.86 & -89.35& 0.65 & 16.26&  0.86 & -15.41&  0.43 &   (1;1;21) \\
123 &     SW Lac &     216598 &     113052 & 22 53 41.66 & +37 56 18.61 & 88.16 & 1.25 &  10.79& 0.82 & 12.30&  1.26 & -14.46&  0.90 &   (1;1;62) \\
124 &     AB And &            &     114508 & 23 11 32.09 & +36 53 35.11 & 109.02& 1.55 & -53.56& 0.94 &  8.34&  1.48 & -27.53&  0.67 &   (1;1;21) \\
125 &   V351 Peg &     220659 &     115627 & 23 25 25.19 & +15 41 19.14 & 17.43 & 0.90 & -16.33& 0.68 &  7.34&  0.92 &  -8.08&  0.89 &   (1;1;14) \\
126 &     VZ Psc &            &     115819 & 23 27 48.39 & +04 51 23.96 & 435.01& 2.46 & 180.24& 1.66 & 16.77&  2.07 &  -4.30&  1.80 &   (1;1;63) \\
127 & BD +14 5016&            &            & 23 36 55.37 & +15 48 06.43 & 19.50 & 1.90 &   2.60& 1.80 &  3.06&  0.38 &   4.20&  0.80 &   (4;5;64) \\
128 &     EL Aqr &            &     117317 & 23 47 18.35 &$-$08 05 12.09& -7.23 & 2.53 &   1.00& 1.65 &  4.71&  1.89 &  12.51&  2.08 &   (1;1;14) \\
129 &      U Peg &            &            & 23 57 58.48 & +15 57 10.08 & -46.50& 1.19 & -53.88& 0.64 &  7.18&  1.43 & -28.50&  1.60 &   (1;1;65) \\
\hline
\end{tabular}
(1) Perryman et al. 1997, (2) Hog et al. 2000, (3) Roese \& Bastian 1988, (4) Hog et al. 1998, (5) Rucinski 2004, (6) Hilditch \& King 
1986, (7) Metcalfe 1999, (8) Rucinski et al. 2003, (9) Maceroni 1986, (10) Milone et al. 1985b, (11) Lu \& Rucinski 1999, (12) Duerbeck 1978, 
(13) Struve et al. 1950, (14) Rucinski et al. 2001, (15) Milone et al. 1995, (16) Lu 1991, (17) Rucinski et al. 2002, (18) Rucinski et al. 2000, 
(19) Yamasaki et al. 1988, (20) Nesci et al. 1986, (21) Pych et al. 2004, (22) Przybylski \& Morris 1965, (23) Goecking et al. 1994, 
(24) Hilditch et al. 1992, (25) Lu 1993, (26) Rucinski \& Lu 1999, (27) Struve 1950, (28) McLean \& Hilditch 1983, (29) Whelan et al. 1979, 
(30) King \& Hilditch 1984, (31) Joy 1926, (32) Rucinski et al. 1993, (33) Lu 1988a, (34) Hrivnak et al. 1984, (35) Lu et al. 2001, 
(36) Hrivnak 1993, (37) Moore \& Paddock 1950, (38) McLean 1981, (39) Hilditch et al. 1989, (40) Sistero \& Sistero 1974,
(41) Hill \& Barnes 1972, (42) Rucinski et al. 1977, (43) Lu \& Rucinski 1993, (44) Milone et al. 1985a, (45) Lu 1988b, (46) Cerruti \& Niemela 1982, 
(47) Rainger et al. 1990b, (48) Rainger et al. 1990a, (49) Batten \& Lu 1986, (50) Sanford 1934, (51) Nelson et al. 1995, (52) Schoeffel 1979, 
(53) McLean 1983, (54) Lu 1986, (55) Pazhouhesh \& Edalati 2003, (56) Goecking \& Duerbeck 1993, (57) Hrivnak 1989, (58) Bell et al. 1990, 
(59) Hill 1989, (60) Ahn et al. 1992, (61) Samec \& Hube 1991, (62) Zhai \& Lu 1989, (63) Hrivnak \& Milone 1989, (64) Maciejewski et al. 2003, 
(65) Lu 1985, (66) Hill et al. 1989, (67) Montgomery et al. 1993, (68) Gatewood \& de Jonge 1994, (69) Gatewood 1995  
          
}
\end{table*} 

\subsection {Parallaxes and proper motions}

The parallaxes and the proper motions in Table 1 were mostly taken from the 
Hipparcos and Tycho Catalogs (ESA\ 1997) and the Tycho Reference Catalog 
(Hog et al.\ 1998). The parallax and the proper motion components of three stars 
(TX Cnc, EE Cet, BL Eri) were taken from Roese \& Bastian\ (1988). Among the 129 
systems in Table 1, the 25 binaries do not have any trigonometric parallax, and 
the 12 binaries have a large relative uncertainty ($\sigma_{\pi}/\pi>0.5$) in 
the parallax measurements. Considering the fact that Hipparcos measurements are 
reliable up to 500 pc (two-sigma detection limit $\sigma \approx 1$ mas, Perryman et 
al.\ 1997), we decided to discard the parallaxes of five (AQ Tuc, TY Pub, XZ Leo, 
AG Vir, RZ Com) systems since their measured parallaxes indicate a distance 
further than 500 pc. These five stars and ER Ori, which were erroneously listed 
with a negative parallax, are treated like the other 25 systems without a 
trigonometric parallax. For them, the photometric parallaxes, which were 
determined from the period-color-brightness relation 
$M(V)=-4.44\log P+3.02(B-V)_{o}+0.12$ by Rucinski\ (2004), were used. The 
photometric parallaxes were calculated with the corrections of the interstellar 
absorption, which were computed from the color excess as $A_{V}=3.1E(B-V)$. Since 
all systems were assigned a spectral type, the color excesses were available 
together with their observed colors.

The Hipparcos and Tycho catalogues usually give an associated uncertainty for all 
of the parallax and the proper motion component measurements. However, there are 
18 systems in the list without an uncertainty at the proper motion components. 
With an optimistic approach, we have preferred to adopt an average uncertainty 
2.5 mas yr$^{-1}$ for them, which were determined by Hog et al.\ (2000) for the 
Hipparcos stars in general. On the other hand, Rucinski\ (2004) estimated that the 
absolute magnitudes computed by the period-brightness relation of W UMa systems are 
within the accuracy of $\pm$0.25 mag. Therefore, a corresponding 12.5 per cent 
uncertainty for the photometric parallaxes were taken.

Recent investigations, both visual and spectroscopic, indicated the presence of 
additional components for some W UMa binaries. Trigonometric parallaxes are frequently 
wrong for those systems as for the brighter systems with large uncertainties in the 
parallaxes. We have not preferred to introduce additional uncertainties into the 
published data. However, for the reader is who want to estimate the bias introduced by 
such multiple system, the W UMa's with additional components are marked in the last 
column of Table 5 and are shown by empty symbols in the figures.    

\subsection {Radial velocities}

All W UMa binaries have circular orbits. Thus the center of mass velocity can 
be obtained simply by fitting a sine curve to the measured radial velocities. 
However, except for the four systems (TW Cet, RZ Tau, TY Pup, AK Her), all W UMa 
stars in our list had well determined orbital parameters so that the center of 
mass velocities ($\gamma$) were available with uncertainties. 
The center of mass velocities and the associated standard errors were listed in 
Table 1 with their sources. If there was more than one source available, the 
data in the latest source is taken since previous radial velocities were used in 
the later determined orbital parameters. Nevertheless, we have found the three 
systems (V523 Cas, AG Vir, V566 Oph) with multiple orbits which were determined 
independently from the independent data sets. For them, the weighted mean of the 
systemic velocities ($\gamma$) and the weighted mean of the associated errors are 
adopted. Only the four systems (TW Cet, RZ Tau, TY Pup, AK Her) had measured center 
of mass velocities but uncertainties were not available. Their uncertainties were 
estimated from the scatter of component velocities on the radial velocity 
curve. All different types of errors have been transformed to standard errors 
for the sake of consistency. 

\section {Galactic space and velocity distributions}

\subsection{Sky and space distributions}

There exist no program similar to that of DDO, which contributed 54 per cent of radial 
velocities, in the southern hemisphere, where the number of observatories already 
relatively less so there is a strong bias in the sky distribution of the present 
W UMa systems. The celestial coordinates ($\alpha, \delta$) are plotted in Fig. 1a that 
the bias is obvious as there are 103 systems (80 per cent) in the northern and 26 systems 
(20 per cent) in the southern hemisphere. Unfortunately, the bias is unavoidable because 1) 
The present work is established on the accumulated data in the literature and it is not vise 
to wait for another unknown period without a clue how southern systems will be 
increased. 2) Eliminating some northern systems to establish isotropy is not feasible since 
it would greatly reduce the statistical significance of the sample. Due to the angle 
(63$\degr$) between the celestial equator and the galactic plane, distribution of the 
sample in the galactic coordinates (Fig. 1b) appears to be less effected by the bias.  
Nevertheless, this is only an illusion because un-isotropy would not change by changing 
the coordinate system.         

\begin{figure}
\resizebox{8cm}{9.6cm}{\includegraphics*{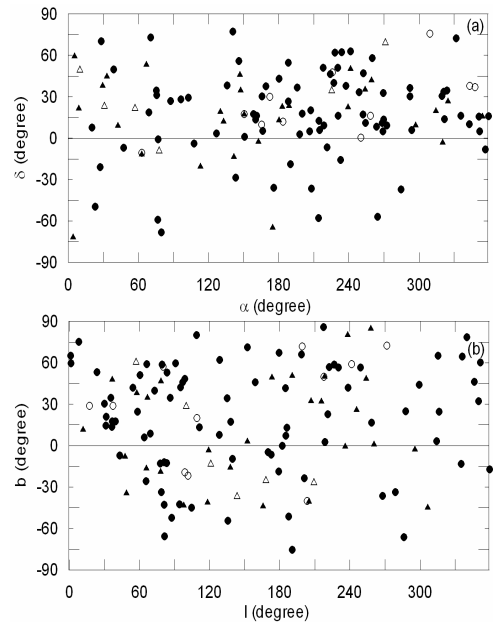}}  
\caption{Sky distributions of the present sample (a) at the celestial, 
(b) at the galactic coordinates. Circles: Hipparcos parallaxes; Triangles: the systems 
use Rucinski's calibration; Filled: normal W UMa; Empty: W UMa's with visible companions.}
\end{figure}

In order to inspect the space distributions of the present W UMa sample, the Sun 
centered rectangular galactic coordinates ($X$ towards Galactic center, 
$Y$ Galactic rotation, $Z$ north Galactic Pole) were calculated. The ($X$, $Y$, $Z$) 
space coordinates are given in Table 2. The projected positions on the Galactic plane 
($X$, $Y$ plane) and on the plane perpendicular to it ($X$, $Z$ plane) are 
displayed in Fig. 2. The bias on the prime reference frames ($X-Y$, $X-Z$, and $Y-Z$) 
are not that obvious as Fig. 1a. Having 69 systems with $X<0$ and 60 system with 
$X>0$, the $Y-Z$ plane shows the least bias. Having 65 and 44 systems the northern 
and southern Galactic latitudes, Galactic plane ($X-Y$) shows the strongest 
bias (Fig. 2) but still it is not obvious as Fig. 1a.

\begin{table*}
\caption{Kinematic data of the W UMas.}
{\small
{\scriptsize
\begin{tabular}{crrrrrrrrrrrrrrrll}
\hline
ID & Name & HD & $l (\degr)$ & $b (\degr)$ &   $X(pc)$ &   $Y(pc)$ &   $Z(pc)$ & \multicolumn{2} {c} {U (km s$^{-1}$)} & \multicolumn{2} {c} {V (km s$^{-1}$)}& \multicolumn{2} {c} {W (km s$^{-1}$)}&   Code &         MG &         OC \\
\hline
         1 &     AQ Tuc &       1372 &     306.66 &     -44.97 &        149 &       -200 &       -249 & -22.76 &   4.77 & -45.28  &  4.28 & -9.93  &  1.43 &             &            &            \\
         2 &     BH Cas &            &     119.08 &      -3.49 &       -144 &        259 &        -18 & 19.39  &  3.76 & -15.02  &  2.79 & -42.25  &  6.37 &            &            &            \\
         3 &     DZ Psc &            &     118.30 &     -41.20 &        -74 &        137 &       -136 & 3.73   & 2.42 & -3.05  &  2.69 & 25.60  &  2.71 &            &            &            \\
         4 &   V523 Cas &            &     121.08 &     -12.59 &        -41 &         67 &        -18 & -43.78 &   5.37 & -35.60  &  4.21 & -28.31 &   3.65 &            &            &            \\
         5 &     AQ Psc &       8152 &     135.65 &     -54.53 &        -52 &         51 &       -101 & 4.38  &  0.70 & 3.94  &  1.57 & 14.81  & 0.81 &            &            &            \\
         6 &     AE Phe &       9528 &     286.22 &     -66.25 &          5 &        -19 &        -45 & -20.12  &  0.84 & -26.19 &   1.53 & 18.91 &   2.32 &       1, 1 & LA, IC &            \\
         7 &     TW Cet &            &     190.59 &     -75.49 &        -25 &         -5 &        -98 & -16.25  &  7.52 & -20.14  &  7.04 & -15.53 & 24.24 &    1, 1, 2 & Cas, IC, LA &            \\
         8 &   V776 Cas &            &     128.21 &       7.81 &       -126 &        160 &         28 & -2.16  &  4.33 & -30.62  &  3.90 & 5.20  &  2.95 &            &            &            \\
         9 &     QX And &            &      11.91 &      11.43 &        365 &         77 &         75 & -24.12  &  3.70 & -16.87  &  4.41 & -17.36 &   4.39 &       1, 1 & LA, IC &    NGC 752 \\
        10 &     SS Ari &            &     143.57 &     -35.94 &       -138 &        102 &       -125 & -16.66  &  2.87 & -24.10  &  3.63 & 0.86  &  2.18 &       1, 1 & IC, LA &            \\
        11 &     GZ And &            &     137.72 &     -15.87 &       -116 &        105 &        -44 & 5.46 &   1.90 & 7.72  &  1.89 & -4.07  & 1.95 &          1 &        UMa &            \\
        12 &   V376 And &      15922 &     139.55 &      -9.63 &       -148 &        126 &        -33 & -51.49  &  5.50 & -18.45 &   5.92 & 7.15 &2.03 &            &            &            \\
        13 &     EE Cet &      17613 &     165.58 &     -43.91 &        -96 &         25 &        -95 & -19.66 &   3.26 & -19.71 &   4.03 & 11.64 &   2.87 &            &            &            \\
        14 &     UX Eri &            &     187.70 &     -51.44 &        -94 &        -13 &       -119 & -15.59 &   3.79 & -14.45 &   6.13 & -2.20 &   3.67 &       1, 1 & IC, LA &            \\
        15 &     EQ Tau &            &     168.13 &     -24.65 &       -128 &         27 &        -60 & -81.50 &   2.42 & -31.50 &   5.71 & -14.52 &   2.48 &            &            &            \\
        16 &     BL Eri &            &     205.10 &     -40.70 &       -248 &       -116 &       -235 & -32.39 &   3.65 & -23.90 &   4.63 & -11.85 &   4.04 &            &            &            \\
        17 &     YY Eri &      26609 &     203.50 &     -40.04 &        -39 &        -17 &        -36 & 40.04 &   2.22 & 2.37 &   0.57 & -20.96  &  2.28 &            &            &            \\
        18 &     AO Cam &            &     152.16 &       2.94 &        -96 &         51 &          6 & 3.61 &   1.25 & -18.18 &   1.92 & -3.46  &  1.33 &            &            &            \\
        19 &     RZ Tau &     285892 &     179.21 &     -18.72 &       -165 &          2 &        -56 & 1.85 & 47.40 & 9.78  &  3.31 & -20.82 & 17.25 &            &            &            \\
        20 &     DN Cam &      29213 &     137.97 &      17.30 &       -158 &        142 &         66 & -12.77 &   1.22 & 9.48  &  1.34 & -20.48  &  4.48 &          1 &        UMa &            \\
        21 &   V410 Aur &     280332 &     170.06 &      -4.65 &       -206 &         36 &        -17 & -39.23 &   3.88 & -2.64 & 11.36 & 0.87 &   6.13 &            &            &            \\
        22 &   V402 Aur &     282719 &     172.79 &      -6.45 &       -141 &         18 &        -16 & -41.93  &  0.96 & -1.58  &  1.56 & -1.78  &  1.09 &            &            &            \\
        23 &     AP Dor &      33474 &     267.96 &     -36.37 &         -6 &       -173 &       -127 & -82.31 & 17.16 & -30.34 &   5.03 & 33.94   & 7.58 &            &            &            \\
        24 &  V1363 Ori &     289949 &     201.09 &     -23.44 &        -90 &        -35 &        -42 & -22.53 &   2.85 & -34.53 &   5.65 & -15.88 &   1.27 &            &            &            \\
        25 &     ER Ori &            &     209.20 &     -26.13 &       -126 &        -70 &        -71 & -16.64 &   2.97 & -32.93 &   2.67 & -20.00  &  1.87 &            &            &            \\
        26 &     RW Dor &     269320 &     278.77 &     -33.62 &         14 &        -92 &        -62 & -23.25 &   7.63 & -85.23  &  6.89 & 0.64 &   8.40 &            &            &            \\
        27 &   V781 Tau &     248087 &     182.21 &      -0.15 &        -81 &         -3 &          0 & -23.71 &   1.30 & -12.75  &  1.37 & -43.84 &   4.83 &            &            &            \\
        28 &     AH Aur &     256902 &     185.15 &       7.31 &       -160 &        -14 &         21 & -29.16 &   1.60 & -16.32  &  4.75 & 10.01  &  2.54 &            &            &            \\
        29 &     QW Gem &     264672 &     186.19 &      12.82 &       -237 &        -26 &         54 & 6.50 &   7.06 & -30.72 & 41.87 & 6.47 & 12.26 &            &            &            \\
        30 &   V753 Mon &      54975 &     218.61 &       2.56 &       -149 &       -119 &          9 & -26.33 &   0.78 & -25.43 &   0.81 & -12.98  &  3.04 &          2 &    IC&            \\
        31 &     TY Pup &      60265 &     236.04 &      -0.72 &       -120 &       -178 &         -3 & -51.62 & 17.50 & 4.56 & 25.11 & -1.09  & 1.11 &            &            &            \\
        32 &     FG Hya &            &     221.03 &      22.83 &       -238 &       -207 &        133 & 61.63 & 33.93 & -78.64 & 53.82 & -37.89&23.38 &            &            &            \\
        33 &     TX Cnc &            &     206.62 &      32.17 &       -145 &        -73 &        102 & -42.09 &   3.67 & -22.51 & 3.60 & -24.46& 5.27 &            &            &       M 44 \\
        34 &     AH Cnc &            &     215.68 &      31.99 &       -574 &       -412 &        441 & -6.11  &  7.33 & -35.39  &  9.38 & -28.54 &   9.59 &            &            &       M 67 \\
        35 &     UV Lyn &            &     184.74 &      41.51 &        -91 &         -8 &         81 & -27.58  &  5.69 & 12.21 &   2.45 & -30.62&    6.15 &            &            &            \\
        36 &     FN Cam &      79886 &     135.15 &      34.44 &       -162 &        161 &        156 & -30.57 &   5.05 & -22.65 &   8.09 & 19.53 &   3.45 &          1 &         LA &            \\
        37 &     EZ Hya &            &     245.81 &      25.69 &        -68 &       -151 &         80 & -20.97  &  3.38 & -14.28 &   1.31 & -11.86 &   3.29 &       1, 2 & LA, IC &            \\
        38 &      S Ant &      82610 &     258.55 &      16.61 &        -14 &        -71 &         21 & -30.51  &  1.82 & 8.15 &   1.70 & -12.40 &   0.80 &            &            &            \\
        39 &      W UMa &      83950 &     158.92 &      45.90 &        -32 &         12 &         36 & 21.01  &  0.82 & -13.19 &   0.45 & -17.29 &   0.91 &            &            &            \\
        40 &     AA UMa &            &     173.25 &      49.21 &       -203 &         24 &        237 & 20.99  &  3.09 & -9.41 &   3.81 & -26.42  &  2.82 &            &            &            \\
        41 &     RT Lmi &            &     190.74 &      50.64 &       -203 &        -38 &        252 & 13.20  &  3.29 & -6.83  &  3.98 & -4.65   & 2.82 &            &            &            \\
        42 &     XY Leo &            &     217.80 &      49.74 &        -32 &        -25 &         48 & 45.60 &  2.46 & 6.51  &  1.72 & -32.13 &   1.70 &            &            &            \\
        43 &     XZ Leo &            &     218.44 &      49.80 &       -195 &       -155 &        295 & -19.18 &   3.99 & 0.06  &  1.54 & -18.35 &   3.03 &       1, 1 & LA, Cas &            \\
        44 &      Y Sex &      87079 &     238.49 &      41.94 &        -43 &        -70 &         74 & -0.95  &  1.88 & -7.13 &   1.58 & 6.14  &  1.87 &            &            &            \\
        45 &     ET Leo &      91386 &     222.34 &      56.81 &        -29 &        -27 &         60 & -17.73 &   1.15 & -13.03  &  0.67 & 10.95&    1.01 &            &            &            \\
        46 &     UZ Leo &            &     230.28 &      56.60 &        -56 &        -68 &        133 & -8.66 &   3.55 & -1.22 &   1.46 & -14.91 &   2.32 &          1 &        UMa &            \\
        47 &     EX Leo &      93077 &     226.62 &      58.85 &        -36 &        -38 &         87 & 1.33  &  0.77 & -15.00  &  2.24 & -19.42&    1.49 &            &            &            \\
        48 &     VY Sex &      93917 &     253.81 &      48.30 &        -25 &        -87 &        102 & 15.60  &  2.67 & -56.00  &  5.65 & -21.28 &   4.39 &            &            &            \\
        49 &     AM Leo &            &     241.50 &      59.02 &        -19 &        -35 &         66 & 1.26  &  1.62 & -12.59 &   3.08 & -0.05  &  1.87 &          1 &     Cas &            \\
        50 &     VW Lmi &      95660 &     198.56 &      66.04 &        -48 &        -16 &        114 & 5.48 &   0.98 & -0.49 &   0.42 & 7.51 &   0.80 &       1, 2 & Cas, UMa &            \\
        51 &     AP Leo &            &     249.13 &      56.47 &        -24 &        -62 &        101 & 67.11&  12.50 & 2.03&    2.59 & -13.45 &   2.39 &            &            &            \\
        52 &     HN UMa &            &     179.28 &      67.35 &        -66 &          1 &        159 & 63.15&  11.73 & -11.70 &   2.90 & -13.80&    4.94 &            &            &            \\
        53 &     AW UMa &      99946 &     199.11 &      71.93 &        -19 &         -7 &         63 & 7.01  &  0.54 & -65.57  &  4.02 & -22.78 &   1.55 &            &            &            \\
        54 &     TV Mus &     310730 &     295.42 &      -2.99 &         99 &       -209 &        -12 & 34.61 &   6.06 & 10.03  &  2.97 & 0.76 &   3.72 &            &            &            \\
        55 &   V752 Cen &     101799 &     287.63 &      25.01 &         29 &        -91 &         44 & -8.12 &   2.97 & -39.68  &  2.37 & -5.22 &   2.77 &            &            &            \\
        56 &     AG Vir &     104350 &     260.60 &       0.72 &        -33 &       -197 &          3 & 11.70  &  1.43 & -13.09  &  1.99 & -10.20&    0.94 &            &            &            \\
        57 &     HX UMa &     104425 &     152.16 &      71.14 &        -43 &         23 &        142 & 51.86 & 21.15 & 4.70 &   3.77 & -5.90 &   5.94 &            &            &            \\
        58 &     CC Com &            &     238.04 &      79.97 &         -7 &        -12 &         79 & -41.89 &   5.34 & -13.52 &  2.07 & -16.34 &   1.51 &            &            &            \\
        59 &     AH Vir &     106400 &     271.52 &      72.39 &          1 &        -28 &         88 & 42.13&  11.89 & -30.79 &   8.29 & -3.19  &  2.86 &            &            &   Wolf 630 \\
        60 &     II UMa &     109247 &     128.65 &      62.13 &        -58 &         72 &        175 & -29.02&  10.65 & -32.09 & 10.61 & -4.69 &   1.45 &       1, 1 & Hya, IC &            \\
        61 &     RW Com &            &     217.61 &      85.87 &         -5 &         -4 &         87 & -33.71 &   7.99 & -37.02 &   8.48 & -56.70 &   1.41 &            &            &            \\
        62 &     RZ Com &            &     257.75 &      84.70 &         -4 &        -19 &        212 & 21.35 &   3.06 & -0.24 &   1.39 & -1.42  &  1.67 &            &            &       Coma \\
        63 &     SX Crv &     110139 &     299.25 &      43.99 &         32 &        -58 &         64 & 19.62 &   1.76 & 1.69  &  1.06 & 5.01 &   0.72 &            &            &            \\
        64 &     BI CVn &            &     108.86 &      80.19 &        -10 &         30 &        186 & 3.13  &  1.55 & -16.11  &  4.86 & -2.52   & 1.97 &          1 &     Cas &            \\
        65 &     KZ Vir &     114726 &     315.40 &      65.02 &         46 &        -45 &        139 & -5.49 &   0.99 & -1.82  &  0.67 & -3.29  &  0.48 &       1, 1 & UMa, Cas &            \\
 \hline
\end{tabular}  
}
}
\end{table*}

\begin{table*}
\contcaption{}
{\small
{\scriptsize
\begin{tabular}{crrrrrrrrrrrrrrll}
\hline
ID & Name & HD & $l (\degr)$ & $b (\degr)$ &   $X(pc)$ &   $Y(pc)$ &   $Z(pc)$ & \multicolumn{2} {c} {U (km s$^{-1}$)} & \multicolumn{2} {c} {V (km s$^{-1}$)}& \multicolumn{2} {c} {W (km s$^{-1}$)}&   Code &         MG &         OC \\
\hline
        66 &     KR Com &     115955 &     339.81 &      78.52 &         14 &         -5 &         75 & 13.68  &  1.03 & -24.99  &  1.71 & -12.35 &   0.49 &            &            &            \\
        67 &     HT Vir &     119931 &     335.97 &      64.42 &         26 &        -11 &         59 & -32.25 &   4.21 & -19.12 &   4.16 & -15.40  &  1.21 &            &            &            \\
        68 &     XY Boo &            &       8.30 &      75.08 &         87 &         13 &        329 & -54.47 & 31.71 & 10.42 &   6.48 & 16.05 &   8.36 &            &            &            \\
        69 &   V757 Cen &     120734 &     316.25 &      24.68 &         46 &        -44 &         29 & 6.12  &  2.73 & -66.49 &   3.15 & 12.13 &   1.28 &            &            &            \\
        70 &     RR Cen &     124689 &     314.13 &       3.15 &         71 &        -73 &          6 & -28.95  &  1.77 & -8.09 &   1.73 & -2.83 &   0.34 &            &            &            \\
        71 &     VW Boo &            &       1.29 &      65.15 &        122 &          3 &        264 & 54.78 & 24.84 & -46.53 & 25.75 & 1.43 & 11.26 &            &            &            \\
        72 &     NN Vir &     125488 &     351.55 &      60.16 &         52 &         -8 &         91 & -16.14 &   1.68 & -24.65 &   3.03 & 0.02 &   0.89 &          1 &     Cas &            \\
        73 &     EF Boo &     234150 &      90.81 &      59.53 &         -1 &         85 &        144 & 2.98  &  1.23 & -40.40  &  6.03 & 8.54  &  3.66 &            &            &            \\
        74 &     CK Boo &     128141 &       1.46 &      59.62 &         79 &          2 &        135 & 89.49& 14.89 & -13.67 &   3.02 & -9.34   & 8.71 &            &            &            \\
        75 &     GR Vir &     129903 &     346.23 &      46.34 &         36 &         -9 &         38 & -65.96 &   1.31 & 6.30 &   0.48 & -36.32  &  1.19 &            &            &            \\
        76 &     AC Boo &            &      79.26 &      58.76 &         13 &         67 &        113 & -10.53  &  1.55 & -5.11 &   1.32 & -6.05  &  1.94 &       1, 2 & UMa, Cas &            \\
        77 &     TY Boo &            &      57.19 &      61.33 &         45 &         69 &        151 & -60.77  &  6.19 & -32.24  &  2.71 & -15.53 &   3.08 &            &            &            \\
        78 &     44 Boo &     133640 &      80.37 &      57.07 &          1 &          7 &         11 & -17.74  &  0.22 & -26.52  & 0.32 & -2.48 &   0.38 &       1, 1 & LA, IC &            \\
        79 &     TZ Boo &            &      66.21 &      59.00 &         31 &         70 &        127 & -77.39 & 15.00 & -34.97  &  4.05 & -5.34  &  5.93 &            &            &            \\
        80 &     BW Dra &            &      98.92 &      48.20 &         -7 &         43 &         49 & -46.26 & 17.50 & -60.90 &   7.33 & -34.00 &   4.21 &            &            &            \\
        81 &     BV Dra &     135421 &      98.92 &      48.20 &         -7 &         44 &         50 & -49.41  &  9.44 & -63.77 &   4.19 & -32.77 &   2.56 &            &            &            \\
        82 &     FI Boo &     234224 &      83.58 &      52.83 &          7 &         63 &         84 & -33.41  &  6.64 & -20.03 &   1.11 & -20.57 &   1.25 &            &            &            \\
        83 &     OU Ser &     136924 &      23.62 &      53.38 &         32 &         14 &         46 & -90.14  &  3.02 & -85.33 &   3.85 & 6.68  &  3.21 &            &            &            \\
        84 &     VZ Lib &            &     350.03 &      32.15 &        169 &        -30 &        108 & -31.88  &  3.54 & -7.19  &  4.97 & -9.09  &   3.66 &            &            &            \\
        85 &     FU Dra &            &      97.07 &      45.80 &        -14 &        111 &        115 & -104.20 & 17.84 & -132.80&  21.82 & 100.22&  18.93 &            &            &            \\
        86 &     YY CrB &     141990 &      60.49 &      51.14 &         27 &         48 &         69 & -19.64   & 1.34 & -20.46  &  1.48 & 15.69  &  1.66 &          1 &         LA &            \\
        87 &     AU Ser &            &      36.85 &      47.84 &         98 &         74 &        136 & -15.35  &  3.18 & -70.45 &   5.81 & -37.06 &   2.17 &            &            &            \\
        88 &   V842 Her &            &      78.53 &      46.61 &         22 &        111 &        120 & -33.79  &  3.39 & -47.35  &  2.04 & -30.18 &   2.28 &            &            &            \\
        89 &     BX Dra &            &      95.18 &      42.39 &        -20 &        218 &        199 & -18.02  &  4.90 & -24.07 &  3.20 & -13.61  &  2.91 &          2 &         LA &            \\
        90 &   V899 Her &     149684 &      54.48 &      41.85 &         54 &         75 &         83 & 16.82  &  2.58 & -7.59   & 0.89 & -30.70  &  2.08 &            &            &            \\
        91 &   V502 Oph &     150484 &      17.22 &      28.85 &         71 &         22 &         41 & -43.03 &   1.01 & -11.85  &  0.55 & -8.35 &   1.39 &            &            &            \\
        92 &   V918 Her &     151701 &      35.67 &      34.70 &         77 &         55 &         65 & -26.09  &  0.97 & -12.83  &  0.51 & -5.68 &   1.11 &            &            &            \\
        93 &   V921 Her &     152172 &      73.07 &      39.89 &         93 &        307 &        268 & -32.59  &  3.21 & -46.14  &  5.14 & -62.25 &   4.97 &            &            &            \\
        94 &   V829 Her &            &      57.87 &      37.91 &         31 &         49 &         45 & -2.89  &  0.96 & -13.92   & 1.05 & -5.69   & 1.06 &          1 &     Cas &            \\
        95 &  V2357 Oph &            &      30.01 &      30.29 &        139 &         80 &         94 & 41.96&  21.92 & -70.24 & 23.12 & -43.51  &12.77 &            &            &            \\
        96 &     AK Her &     155937 &      37.51 &      28.73 &         66 &         51 &         46 & 6.27 & 11.35 & -13.70  &  8.24 & -23.09   & 8.52 &            &            &            \\
        97 &   V728 Her &            &      66.80 &      34.53 &        135 &        314 &        235 & 18.32  &  5.03 & 3.22  &  4.24 & 40.57  &  4.80 &            &            &            \\
        98 &     GM Dra &     238677 &      86.44 &      34.70 &          5 &         81 &         56 & 29.12  &  2.72 & 25.10  &  2.05 & -22.99 &   2.65 &            &            &            \\
        99 &  V2377 Oph &     159356 &      31.50 &      20.92 &         79 &         48 &         35 & -5.37 &   2.03 & -35.08  &  2.72 & -15.51 &   0.88 &            &            &            \\
       100 &   V535 Ara &     159441 &     335.18 &     -13.20 &        100 &        -46 &        -26 & -27.84 &   1.48 & -34.51  &  4.23 & 25.82  &  2.26 &            &            &            \\
       101 &  V2388 Oph &     163151 &      36.62 &      17.68 &         52 &         39 &         21 & 13.72  &  1.95 & -60.46 &   2.53 & -9.47 &   0.31 &            &            &            \\
       102 &   V566 Oph &     163611 &      31.17 &      14.40 &         59 &         36 &         18 & -51.35  &  1.51 & 5.19 &   2.22 & -15.60  &  0.60 &            &            &            \\
       103 &   V972 Her &     164078 &      58.48 &      24.76 &         29 &         48 &         26 & 8.36  &  0.38 & -5.23  &  0.48 & 8.10  &  0.34 &            &            &            \\
       104 &   V508 Oph &            &      39.37 &      17.66 &         96 &         79 &         40 & -9.74  &  5.59 & -44.05 &   6.25 & -14.24 &   1.52 &            &            &            \\
       105 &     EF Dra &            &     100.05 &      29.28 &        -27 &        150 &         85 & 23.06  &  3.21 & -38.19 &  3.22 & -10.67 &   2.85 &            &            &            \\
       106 &   V839 Oph &     166231 &      36.43 &      13.47 &         97 &         71 &         29 & -51.27 &   0.46 & -44.85  &  1.50 & 2.60  &  3.19 &            &            &            \\
       107 &    eps Cra &     175813 &     359.54 &     -17.32 &         29 &          0 &         -9 & 58.62  &  1.15 & -21.94  &  0.60 & -5.98   & 0.51 &            &            &            \\
       108 &   V401 Cyg &            &      64.13 &       6.00 &         38 &         77 &          9 & 13.96   & 4.34 & 20.41 &  2.85 & 2.04 &  3.37 &            &          &            \\
       109 &  V2082 Cyg &     183752 &      69.40 &       8.70 &         32 &         84 &         14 & -52.31 &  1.96 & -19.09 &  0.84 & 5.75  &  0.60 &            &            &            \\
       110 &   V417 Aql &            &      43.21 &      -7.03 &         95 &         89 &        -16 & -3.73   & 3.06 & -20.52 &   3.50 & -4.37 &   2.48 &            &            &            \\
       111 &     OO Aql &     187183 &      47.82 &      -8.17 &        102 &        113 &        -22 & -50.97   & 2.41 & -25.70  &  1.39 & -37.60 &  5.46 &            &            &            \\
       112 &     VW Cep &     197433 &     109.22 &      20.06 &         -9 &         25 &          9 & -74.79   & 2.08 & -34.81 &  1.16 & 0.20   & 0.38 &            &            &            \\
       113 &     LS Del &     199497 &      66.00 &     -16.49 &         31 &         69 &        -22 & -67.49 &   6.78 & -5.48  &  2.45 & -15.62 &   2.89 &            &            &            \\
       114 &  V2150 Cyg &     202924 &      77.81 &     -13.04 &         43 &        201 &        -48 & -5.65 &   1.37 & -14.26  &  0.83 & -3.11 &   2.30 &            &            &            \\
       115 &     HV Aqr &            &      48.99 &     -34.47 &         77 &         89 &        -81 & 54.48  &  7.28 & -50.34  &  5.94 & 5.91  &  1.35 &            &            &            \\
       116 &  V1073 Cyg &     204038 &      81.13 &     -11.95 &         28 &        178 &        -38 & 10.46  &  3.45 & -5.33  &  1.43 & -9.53  &  2.73 &            &            &            \\
       117 &     KP Peg &     204215 &      65.75 &     -25.73 &         85 &        188 &        -99 & 5.53 &  3 .88 & -2.13  &  2.72 & -7.30  &  2.74 &       1, 2 & Cas, UMa &            \\
       118 &     DK Cyg &            &      83.29 &     -12.76 &         26 &        219 &        -50 & 29.62 & 14.63 & -8.34  &  1.87 & 7.08 &   2.64 &            &            &            \\
       119 &     BX Peg &            &      78.21 &     -19.00 &         32 &        151 &        -53 & 2.52 &   2.00 & 19.22  &  2.91 & -17.99 &   2.40 &            &            &            \\
       120 &   V445 Cep &     210431 &     111.13 &      13.38 &        -39 &        101 &         26 & -63.88  &  2.65 & 20.92 &   1.25 & 1.51 &  0.55 &            &            &            \\
       121 &     BB Peg &            &      78.90 &     -33.56 &         53 &        271 &       -183 & -15.44  &  4.58 & -50.75 &20.31 & -24.94 & 30.77 &            &            &            \\
       122 &   V335 Peg &     216417 &      81.25 &     -42.83 &          7 &         45 &        -42 & -30.97  &  1.50 & -41.28 &   1.63 & -26.21 &   1.97 &          1 &     Hya &            \\
       123 &     SW Lac &     216598 &      98.68 &     -19.33 &        -12 &         76 &        -27 & -31.07  &  3.22 & -22.24  &  1.24 & -6.60 &   1.25 &       1, 1 & LA, IC &            \\
       124 &     AB And &            &     101.60 &     -21.81 &        -22 &        109 &        -45 & -37.31 &   7.05 & -53.16 &   5.05 & -38.91&   8.74 &            &            &            \\
       125 &   V351 Peg &     220659 &      94.40 &     -42.37 &         -8 &        100 &        -92 & -6.95  &  0.81 & -16.05  &  1.46 & -5.22 &   1.51 &          1 &     IC &            \\
       126 &     VZ Psc &            &      87.50 &     -52.17 &          2 &         37 &        -47 & -133.25 & 16.32 & -12.75  &  1.72 & -8.92 &   2.11 &            &            &            \\
       127 &BD +14 5016 &            &      97.90 &     -43.42 &        -33 &        235 &       -225 & -35.15  &  4.56 & -6.46  & 2.37 & -8.78 &   2.24 &          1 &         LA &            \\
       128 &     EL Aqr &            &      81.40 &     -65.65 &         13 &         87 &       -193 & 4.31  &  3.35 & 8.92 &   2.46 & -9.27 &   2.24 &            &            &            \\
       129 &      U Peg &            &     104.61 &     -45.01 &        -25 &         95 &        -99 & 45.41 &   8.58 & -26.60 &   1.85 & 2.41  &  3.73 &            &            &            \\
\hline
\end{tabular}  
}
}
\end{table*}

\begin{figure}
\resizebox{8cm}{14cm}{\includegraphics*{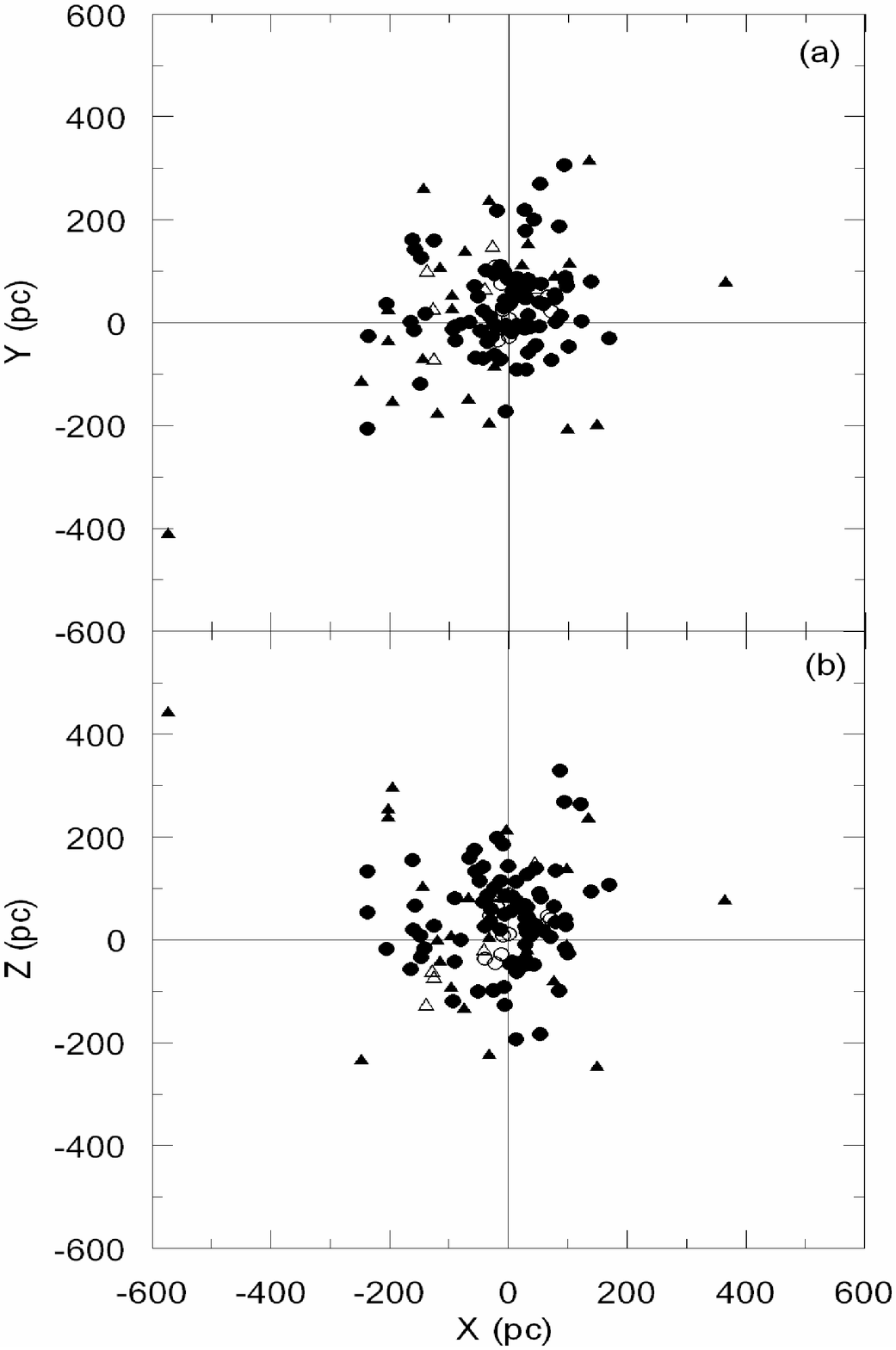}} 
\caption{The space distributions with  respect to the Sun.
$X$, $Y$, and $Z$ are heliocentric galactic coordinates
directed towards the Galactic Center, Galactic rotation and the North Galactic 
Pole. Symbols mean the same as Fig. 1.}
\end{figure}
       
With a median distance of 137 pc, the sample contains relatively nearby systems  
and it can be considered within the Galactic thin disc. 
Occupying almost the same space in the Solar neighborhood, the CAB sample 
(Karata\c{s} et al.\ 2004), had a smaller median distance (98 pc) than W UMa sample. 
Thus, we were curious to compare space dispersions of the present W UMa sample and 
CAB sample. In order to avoid complications of far distant members, the space 
dispersions of both samples within 300 pc are calculated. Indeed, with the space 
dispersions of $\delta X=\pm 78$, $\delta Y=\pm 90$, and $\delta Z=\pm 87$ pc, 
the W UMa sample seems a little more dispersed than the CAB sample with the space 
dispersions of $\delta X=\pm 77$, $\delta Y=\pm 75$, and $\delta Z=\pm 73$ pc. 

Considering the fact that CAB's being on average brighter than the less common 
W UMa's, it has been expected the opposite, since CAB's would be visible at longer 
distances. Obviously, the observational selection is more affected by the easy 
recognition of W UMa's rather than the higher absolute brightness of CABs. Apparently, 
the less bright contact binaries were identified and studied more than the CAB's in 
similar distances.

\subsection{Galactic space velocities}

Galactic space velocity components ($U$, $V$, $W$) were computed by applying 
the algorithm and the transformation matrices of Johnson \& Soderblom\ (1987) 
to the basic data: celestial coordinates ($\alpha$, $\delta$), proper motion 
components ($\mu_{\alpha}$, $\mu_{\delta}$), radial velocity ($\gamma$) and the 
parallax ($\pi$) of each star in Table 1, where the epoch of J2000 coordinates 
were adopted as described in the International Celestial Reference System (ICRS) 
of the Hipparcos and the Tycho Catalogues. The transformation matrices use the 
notation of the right handed system. Therefore, the $U$, $V$, $W$ are the 
components of a velocity vector of a star with respect to the Sun, where the 
$U$ is directed toward the Galactic center ($l=0^{o}, b=0^{o}$); the $V$ is in 
the direction of the galactic rotation ($l=90^{o}, b=0^{o}$); and the $W$ is 
towards the north Galactic pole ($b=90^{o}$). 

Uncertainties of ($U$, $V$, $W$) space velocity components have been computed 
by propagating the uncertainties of the input data (proper motions, parallax 
and radial velocity) by the algorithm also by Johnson \& Soderblom\ (1987). 
The arithmetic mean of the uncertainties for ($U$, $V$, $W$) are: 
$\delta U=\pm 5.52$, $\delta V=\pm 4.67$ and $\delta W=\pm 3.75$ km s$^{-1}$. By 
inspecting the individual errors of space velocity vectors 
($S=\sqrt{U^{2}+V^{2}+W^{2}}$), we have found only 11 (8.5 per cent) systems with 
the errors bigger than $\pm 20$ km s$^{-1}$. If those systems were removed from 
the sample, the averages would reduce to $\delta U=\pm 4.04$, $\delta V=\pm 3.12$, 
and $\delta W=\pm 2.68$ km s$^{-1}$. Thus, it can be claimed confidently that most 
of the sample stars have the velocity uncertainties smaller than the velocity 
dispersions calculated. However, in comparison with the CAB sample 
(Karata\c{s} et al.\ 2004), the average uncertainties of the present 
W UMa sample appear nearly two times bigger. Obviously, this is the result of the 
lower accuracy of W UMa radial velocities because of shorter orbital periods and 
higher rotation rates.  

Having relatively less accurate space velocities, we were curious to check the 
effect of the differential Galactic rotation. Thus, the 
first-order Galactic differential rotation contributions to the $U$ and $V$ 
components were computed as described in Mihalas \& Binney\ (1981). The $W$ 
velocities are not affected in this first order approximation. We have found 27 
stars (21 per cent) with the contribution from the Galactic differential rotation 
being greater than the uncertainty of the $U$ component of the space velocity. 
The effect on the $V$ component is rather small that the effect of Galactic 
differential rotation was found smaller than the uncertainties in $V$. 
Nevertheless, the Galactic differential rotation correction was applied to all 
of stars in the present sample. The corrected $U$, $V$, $W$ are given in 
Table 2, together with the propagated standard errors.  

\begin{figure}
\resizebox{8cm}{14cm}{\includegraphics*{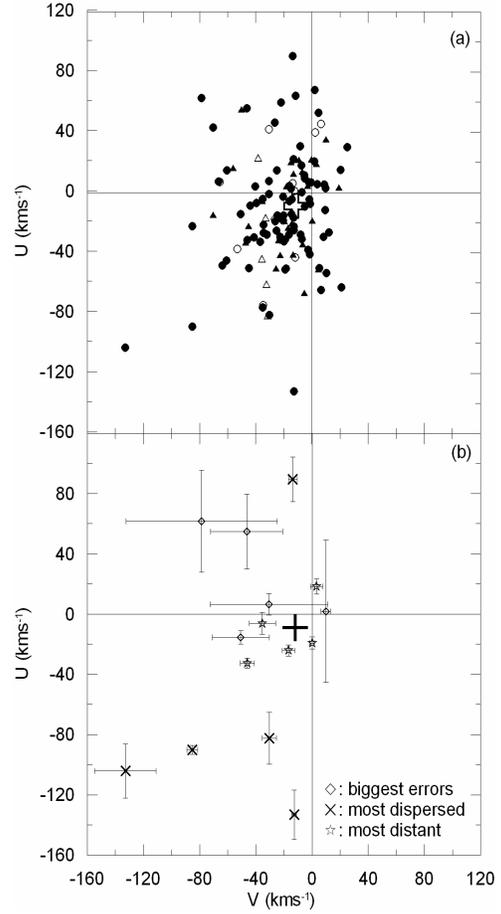}} 
\caption{(a) Velocity dispersions on the $U-V$ plane. 
(b) propagated errors of some selected systems. Symbols in (a) are 
the same as Fig. 1 and symbols in (b) are self indicated. The 
velocities are heliocentric. The position of LSR is marked by +.}
\end{figure}

There is no direct way to estimate how the un-isotropic sky distribution (bias 
shown in Fig. 1a) effects the W UMa space velocities. However, since present 
sample contains only nearby systems within the Galactic thin disc and since 
first order Galactic differential rotation correction were applied, the bias 
seems to be minimized perhaps disappeared in the space velocity distribution. 
This is because the velocity distribution present W UMa sample has a great 
resemblance to the velocity distribution of the CAB sample (Karata\c{s} et 
al.\ 2004) which were considered homogeneously and isotropically distributed 
in the solar neighbourhood. Distribution of the space velocity components on 
the $U-V$ plane is displayed in Fig. 3a.    

In order to display reliability of computed $U$, $V$, $W$ velocities, Fig. 3b 
shows error bars some selected systems as 1) five system which are the most 
distant, 2) five systems with the biggest space velocity vector ($S$), 
3) five systems which are the most dispersed on the $U-V$ diagram. It is 
also clear on Fig. 3b that propagated uncertainties, even the extreme cases as 
above, appear smaller than the dispersion thus the space velocities 
($U$, $V$, $W$) and their dispersion can be considered realible. Nevertheless, 
in order to minimize the contribution of the space velocity errors, the 
dispersion were calculated with respect to the LSR rather than arithmetic 
mean of the space velocities. The position of the LSR is obtained by 
subtracting the Sun's velocity  $(U, V, W)_{\odot}=(9,12,7)$ km s$^{-1}$ 
by Mihalas \& Binney\ (1981) from the frame of ($U$, $V$, $W$) velocities. 

\subsection{Population analysis}

As in the CAB sample (Karata\c{s} et al.\ 2004), we have used the parameter 
$f_{K}=(1/300)(1.0u^{2}+2.5v^{2}+3.5w^{2})^{1/2}$ to determine the 
possible metal poor binaries kinematically as suggested by Grenon\ (1987) and 
Bartkevicius, Lazauskaite, \& Bartasiute\ (1999). Here, the $u$, $v$, $w$ 
velocities represent space velocity components with respect to the LSR. 
Statistically, the stars $f_{K}\leq0.35$ belong to the thin disc, the stars with 
$0.35<f_{K}\leq1.00$ belong to the thick disc. The stars with $f_{K}>1$ belong to 
the halo population. 

According to $f_{K}$ analysis, the vast majority (93 per cent) of our sample are 
thin disc stars. The possible thick disc members are found to be 7 per cent  
in the current sample. The thick disc candidates are FU Dra, AP Dor, 
RW Dor, FG Hya, OU Ser, AU Ser, V921 Her, V2357 Oph, and VZ Psc. Spectroscopic 
metal abundances of W UMa stars appear unachievable due to fast rotation and 
blendings in their spectra. Photometric indices are also entirely unreliable 
due to the elevated magnetic activity of the W UMa systems.  Nevertheless, 
photometric metal abundance of BW Dra has been determined by Marsakov \& 
Shevelev\ (1995) ($[m/H]=-0.56$ dex) and recently by Nordstr\"om et al.\ (2004) 
($[m/H]=-0.69$ dex) from the observations of \it Str\"omgren \rm photometry. 
A 10.33 Gyr age is estimated for it by Marsakov \& Shevelev\ (1995) according 
to the revised Yale isochrones ages (Green, Demarque, \& King \ 1987). If those 
metal abundances and isochrone ages are correct, with the $f_{K}$ kinematic 
value of 0.33, BW Dra appears as a star which belongs to the thick disc 
population although it's $f_{K}$ value is little less than the lower limit.  

Karata\c{s} et al.\ (2004) had found 7 per cent of CAB binaries are possible 
members of the thick disc population. It is indeed interesting that the same 
percentage of thick disc W UMa binaries confirmed in this study coincides with 
nearly the same ratio, 6 per cent of solar neighborhood stars in general belongs 
to the thick disc population which were found by Buser, Rong, \& Karaali\ (1999),  
Siegel et al.\ (2002), and Karaali, Bilir, \& Hamzao\u glu\ (2004).     

\begin{figure*}
\resizebox{17.40cm}{9cm}{\includegraphics*{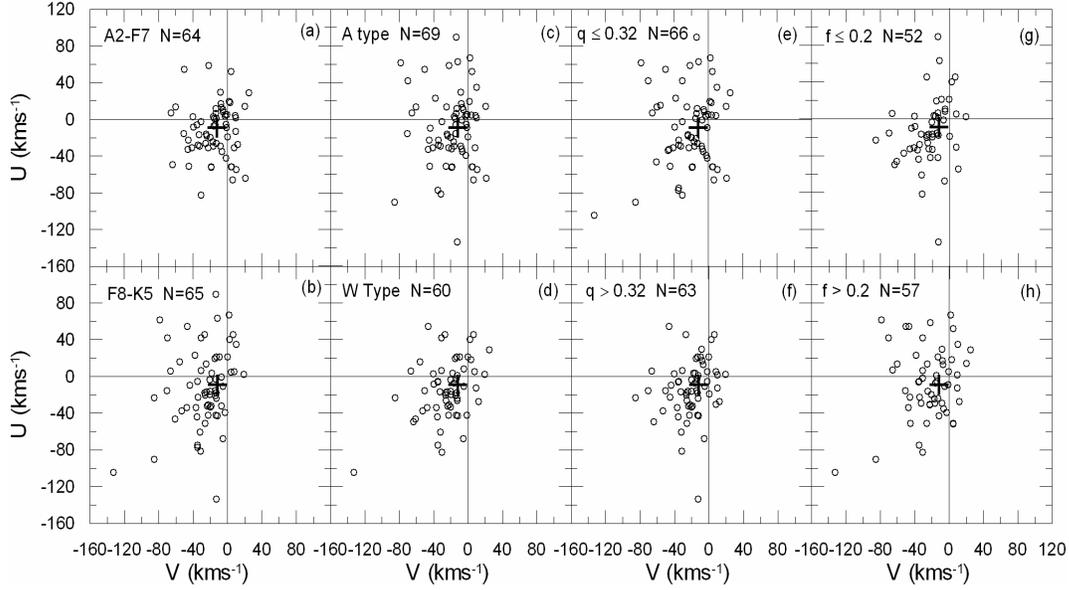}} 
\caption{Comparison of the space velocity dispersions of various sub groups on 
the $U-V$ diagram. (a) earlier spectral types A2 to F7; (b) later spectral types 
F8 to K5; (c) A type; (d) W type; (e) mass ratio $q\leq0.32$; (f) mass ratio 
$q>0.32$; (g) over-contact parameter $f\leq0.2$; (h) over-contact parameter 
$f>0.2$.}
\end{figure*}

\section{Discussion}
\subsection{Kinematics of contact binaries}

Having only the main sequence components from the spectral types A to K, the present 
sample of W UMa binaries appear less heterogeneous than the CAB binaries with the 
components F to M spectral types of all combinations of the luminosity classes from 
V to II. Having orbital periods from 1 days to several hundred days, the CAB sample 
is also considered heterogeneous (Karata\c{s} et al.\ 2004) regarding to the period 
ranges since different orbital periods may represent different evolutionary paths 
(Plavec\ 1968, Thomas\ 1977). But, with a very limited range of short orbital 
periods ($0.22<P<1.13$ days), the W UMa sample can be claimed homogeneous 
in a sense that all binaries with main sequence-components are in the contact stage. 
However, the velocity distribution of W UMa sample on the $U-V$ plane (Fig. 2) 
appears heterogeneous as the CAB sample, that kinematically younger and older 
sub systems occupy the same velocity space. With the dispersions of 36.5, 26.2, 
19.5 km s$^{-1}$ at $U$, $V$, and $W$ velocity components, it is possible to assign 
a 4.43 Gyr kinematical age to the whole sample according to the kinematical tables 
of Wielen\ (1982). Karata\c{s} et al.\ (2004) assigned 3.86 Gyr age to the field 
CAB binaries. Thus, even before selecting out possible moving group (MG) members, 
which indeed seems to exist from the appearance of Fig. 3, the W UMa sample is 
little older than the CAB sample. 

Guinan \& Bradstreet\ (1988) investigated whether correlations exist between 
the space velocities and the physical properties of the W UMa systems, such as 
orbital period ($P$), mass ratio ($q$), system type (A or W), filling factor ($f$), 
ultraviolet excess, and metallicity index. A weak correlation appeared possible in 
the sense that the space velocities tend to increase with decreasing orbital period. 
Also, a weak correlation appeared as binaries with large $f$ (over-contact) tend to 
have higher space velocities. But, no correlation had been found between the space 
velocities and the mass ratio, the space velocities and the binary type (A or W), 
the space velocities and the metallicity index. 

In order to search kinematically different sub groups, here we have preferred to 
divide the whole sample into two sub samples according to a criterion, then to 
compare their $U-V$ diagrams. The preliminary criteria and the $U-V$ diagrams 
are summarized and displayed in Fig. 4.

W UMa binaries with spectral types later than F7 tend to have larger dispersions
(Fig. 4a and b) than W UMa's of earlier spectral types. This could be explained 
by the fact that the later spectral types on the H-R diagram contain older systems 
since evolution into the main-sequence and the duration on the main sequence is longer. 
However, it is indeed extraordinary that kinematically young and old systems (small and large 
dispersions) may well exist in both groups. Thus, this criterion fails to provide 
kinematically homogeneous sub samples.  

W UMa stars are classified into A- or W-type systems from their light curves and 
radial velocity curve (Binnendijk\ 1970). The W-type systems are those like W UMa 
itself in which the hotter component (the star eclipsed at the primary minimum) 
is the smaller and the less massive. Comparison of the $U-V$ diagram between the 
A- and W-type in Fig. 4 confirms the Guinan \& Bradstreet\ (1988) that both groups 
are indistinguishable kinematically, that younger and older systems are equally 
likely to exist in both sub groups of A or W.  
 
If a few high dispersion binaries are removed from the small mass ratio systems, then 
both sub groups (Fig. 4e and f) of mass ratio criterion will be 
indistinguishable kinematically. Like the two criteria before, the mass ratio 
criterion also fails to distinguish younger and older systems homogeneously. 

\begin{figure}
\resizebox{8cm}{7.6cm}{\includegraphics*{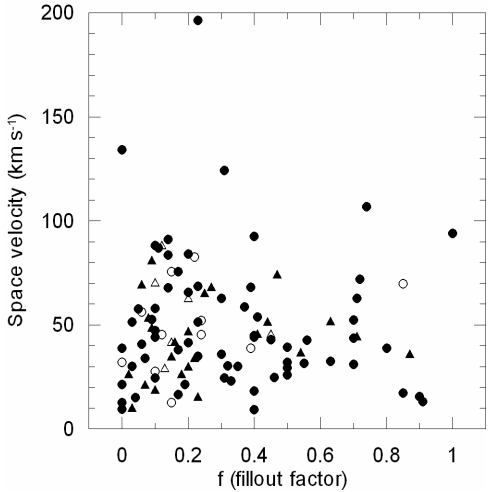}} 
\caption{The space velocities is plotted against the fillout factor ($f$) for 
W UMa binaries.}
\end{figure}

The degree of over-filling ($f$) of the inner critical equipotential surface is an 
important parameter which may indicate a stage of an evolution for a contact binary 
which would coalesce into a single star. FK Com is believed by some to be the product of 
binary star coalescence (Bopp \& Rucinski\ 1981, Bopp \& Stencel\ 1981). The value of 
over-filling factor $f$, which depends on how it is defined, changes from the minimum, 
when the system is just in contact, to the maximum when components fillout their 
outer critical envelopes. Different authors define f differently, e.g.  Guinan \& 
Bradstreet\ (1988) define f as $1 \leq f\leq2$, but Maceroni \& van't Veer\ (1996) 
define f as $0\leq f \leq 1$. Using the same definition as Maceroni \& van't Veer\ 
(1996), we have found 109 systems with ($f$). The systems with smaller $f$ are 
compared to the systems with large $f$ on the $U-V$ diagram (Fig. 4g and h). 
The systems with larger $f$ values seem to have larger dispersions, so they appear 
kinematically older than the systems with smaller $f$. Although this is a natural 
consequence of the theory of coalescence into a single star, and it may appear to 
confirm Guinan \& Bradstreet\ (1988), with a careful look at the $U-V$ diagrams 
(Fig. 4g and h) one may notice the two sub groups are not really homogeneous. 
Kinematically older and younger systems appear to exist in both groups.

In order to make a direct comparison to the Fig. 9 of Guinan \& Bradstreet\ (1988), 
the space velocities ($s=\sqrt{u^{2}+v^{2}+w^{2}}$) in our sample is plotted against 
the $f$ values in Fig. 5. According to this presentation, the high velocity stars may 
well exist at low or high values of $f$. Thus it can be concluded that $f$ criterion, 
like the criteria before (spectral type, binary type (A or W), mass ratio) is not a 
good criterion to differentiate kinematically older and younger W UMa binaries.              

This result, however, is consistent with the very brief lifetime of a contact stage, 
which was predicted by Guinan \& Bradstreet\ (1988) as $0.1<t_{contact}<1$ Gyr, and 
the kinematical ages of the sample as a whole which is 4.43 Gyr. The dispersions on 
the $U-V$ diagram are produced in a time scale not only covering the contact stage but 
also covering the pre-contact stages of all. Obviously the contact stage for many 
W UMa stars is too short to be effective on the observed dispersions. Therefore, it 
is possible that a W UMa system with a longer pre-contact evolution may reach to a 
contact configuration at kinematically larger ages that the $f$ value can be very 
small. But, on the other hand it is also possible for a W UMa system, with a much 
shorter pre-contact stage (small kinematical ages), to have larger value for the $f$. 
Therefore $f$ value criterion fails to discriminate between kinematically young and 
old systems.   

\begin{figure*}
\resizebox{17.40cm}{11.60cm}{\includegraphics*{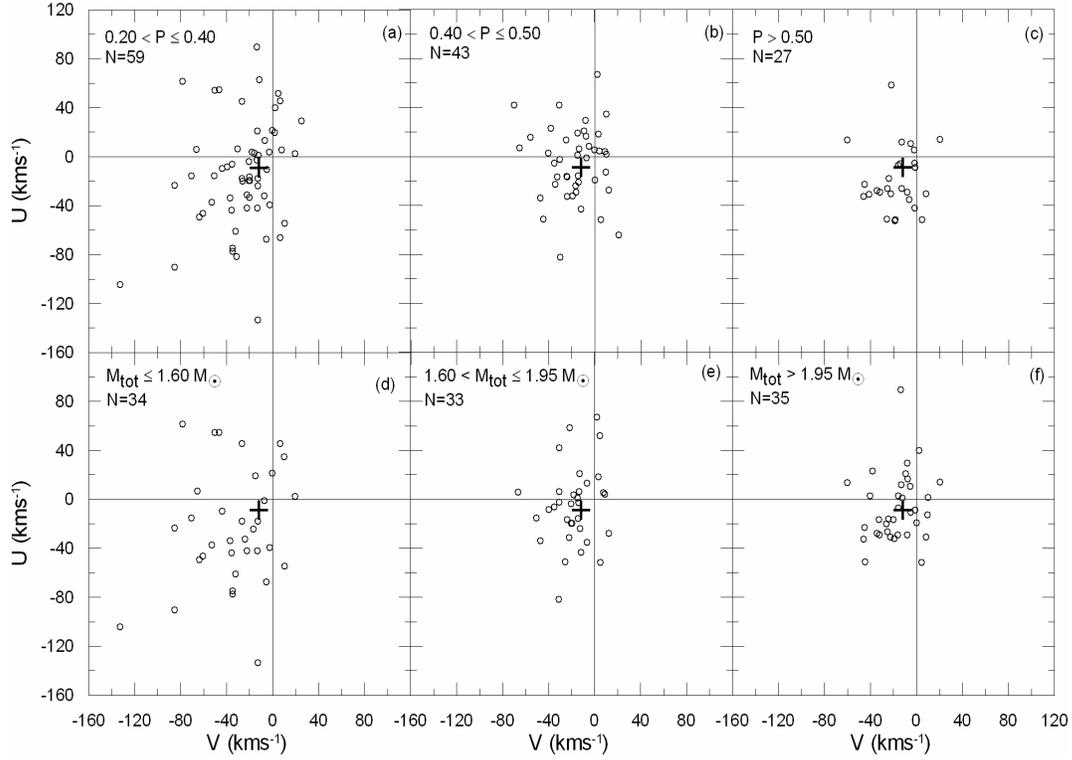}} 
\caption{Comparison of the space velocity dispersions of period and total mass sub 
groups on the $U-V$ diagram. (a) $0.2<P\leq0.4$ days; (b) $0.4<P\leq0.5$ days; 
(c) $P>0.5$ days; (d) $M_{tot}\leq1.60 M_{\odot}$; 
(e) $1.60<M_{tot}\leq1.95 M_{\odot}$; and (f) $M_{tot}>1.95 M_{\odot}$.}
\end{figure*}

The present W UMa sample were said above to be homogeneous in a narrow range 
($0.22<P<1.13$ days) of the short orbital periods. However, the correlations of orbital 
periods with the dispersions on the $U-V$ diagram is well presented in Fig. 6 in 
a sense that the short period W UMa systems have larger dispersions. The sub groups 
according to the total masses ($M=M_{1}+M_{2}$) display a similar correlation 
that less massive systems have larger dispersions than more massive systems. 

All sub groups discussed above are summarized numerically in Table 3. The numerical 
values (velocity averages and dispersions) of the sub groups confirm eye inspection 
of the dispersions on $U-V$ diagrams on the Fig. 4 and Fig. 6. Indeed the grouping 
according to spectral type, systems type (A or W), mass ratio, and $f$ value in 
Fig. 4 do not show very much differences in the dispersions. But grouping according 
to orbital period and total mass (Fig. 6) is confirmed strongly in Table 3. Using 
the tables of Wielen\ (1982), a kinematical age was assigned to each hypothetical 
sub-group. The ages represent a mean kinematical age of all the stars 
in a group. The age to be meaningful depends on the meaningfulness of the 
group. The sub groups using orbital period and total mass criteria are 
indeed meaningful because the decrease of orbital period and total mass, which is 
established for CAB (Karata\c{s} et al.\ 2004), appears to be a process continuing 
also in the contact stage. Because of a relatively very short contact stage, there 
is no guarantee about the sub groups of total mass and orbital period being 
homogeneous. One way of selecting out younger population W UMa binaries among 
the sub samples according to orbital period, is to pick out possible moving 
group members as done in CAB sample by Karata\c{s} et al.\ (2004).     

\setcounter{table}{2}
\begin {table*}
\caption{Kinematical data and the predicted kinematical ages of the sub groups of W UMa.} 
{\scriptsize
\begin{tabular}{lccccccccccccccc}
\hline
Data& N &\multicolumn{2} {c} {$<U>$}&\multicolumn{2} {c} {$<V>$}&\multicolumn{2} {c} {$<W>$}&\multicolumn{2} {c} {$\sigma_{U}$}&\multicolumn{2} {c} {$\sigma_{V}$}&\multicolumn{2} {c} {$\sigma_{W}$}& $\sigma_{tot}$ &  Age\\
   &  &\multicolumn{2} {c} {(km s$^{-1}$)}&\multicolumn{2} {c} {(km s$^{-1}$)}&\multicolumn{2} {c} {(km s$^{-1}$)}&\multicolumn{2} {c} {(km s$^{-1}$)}&\multicolumn{2} {c} {(km s$^{-1}$)}&\multicolumn{2} {c} {(km s$^{-1}$)}& (km s$^{-1}$)&(Gyr)\\
\hline
       All            & 129 & -12.11 & 0.62 & -21.19 & 0.62 & -8.52 & 0.41 & 36.50 & 7.06 & 26.23 & 7.09 & 19.46 & 4.65&49.0 11.0& 4.43 \\
& & & & & & & & & & & & & & & \\
A2-F7 Spectral type   &  64 & -12.28 & 0.96 & -15.22 & 0.51 & -6.16 & 0.55 & 28.94 & 7.68 & 21.34 & 4.11 & 17.22 & 4.43&39.9 ~9.8& 2.83 \\
F8-K5 Spectral type   &  65 & -11.93 & 0.80 & -27.07 & 1.12 & -10.85& 0.60 & 42.98 & 6.42 & 30.55 & 9.07 & 21.46 & 4.86&56.9 12.1 & 5.88 \\
& & & & & & & & & & & & & & & \\
    A Type            &  69 & -10.55 & 1.02 & -18.40 & 0.88 & -7.85 & 0.45 & 39.86 & 8.51 & 24.17 & 7.31 & 15.93 & 3.77&49.3 11.8& 4.48 \\
    W Type            &  60 & -13.90 & 0.62 & -24.40 & 0.89 & -9.29 & 0.71 & 32.21 & 4.80 & 28.41 & 6.89 & 22.85 & 5.48&48.7 10.0& 4.37 \\
& & & & & & & & & & & & & & & \\
  $q\leq0.32$         &  66 & -10.92 & 0.90 & -21.58 & 0.96 & -3.74 & 0.49 & 40.97 & 7.29 & 29.33 & 7.77 & 22.08 & 3.96&55.0 11.6& 5.53 \\
  $q>0.32$            &  63 & -13.34 & 0.86 & -20.78 & 0.80 & -13.54& 0.67 & 31.15 & 6.80 & 22.52 & 6.34 & 16.26 & 5.30&41.7 10.7& 3.14 \\
& & & & & & & & & & & & & & & \\
  $f\leq0.2$         &  52 & -14.06 & 0.73 & -23.16 & 0.31 & -12.24& 0.48 & 36.41 & 5.30 & 24.36 & 2.22 & 18.70 & 3.48&47.6 ~6.7& 4.18 \\
 $f>0.2$             &  57 & -7.10  & 1.24 & -23.33 & 1.41 & -6.43 & 0.82 & 38.17 & 8.93 & 30.32 &10.15 & 21.87 & 5.91&53.4 14.7& 5.24 \\
& & & & & & & & & & & & & & & \\
$0.2<P\leq0.4$        &  59 & -14.02 & 0.95 & -25.82 & 1.22 & -8.57 & 0.79 & 44.00 & 7.33 & 31.64 & 9.37 & 22.15 & 6.06&58.6 13.6& 6.18 \\
$0.4<P\leq0.5$        &  43 & -5.46  & 1.23 & -17.29 & 0.55 & -8.96 & 0.49 & 29.72 & 8.09 & 21.66 & 3.59 & 17.36 & 3.24&40.7 ~9.4& 2.96 \\
$P>0.5$               &  27 & -18.51 & 0.68 & -17.29 & 0.95 & -7.72 & 0.26 & 26.99 & 3.55 & 18.74 & 4.93 & 16.00 & 1.33&36.6 ~6.2& 2.30 \\
& & & & & & & & & & & & & & & \\
$M_{tot}<1.60$        &  34 & -23.26 & 1.44 & -33.87 & 1.70 & -10.65& 0.85 & 48.89 & 8.39 & 39.28 & 9.89 & 26.28 & 4.95&68.0 13.9& 7.87 \\
$1.60<M_{tot}\leq1.95$&  33 & -5.52  & 0.75 & -18.19 & 1.32 & -8.61 & 0.95 & 31.62 & 4.31 & 18.86 & 7.61 & 16.36 & 5.48&40.3 10.3& 2.90 \\
$1.95>M_{tot}$        &  35 & -6.89  & 1.43 & -16.89 & 0.73 & -7.95 & 0.79 & 27.59 & 8.46 & 19.01 & 4.31 & 15.78 & 4.65&37.0 10.6& 2.38 \\
& & & & & & & & & & & & & & & \\
        MG            &  27 & -14.44 & 0.45 & -14.71 & 0.49 & -4.83 & 0.84 & 13.08 & 2.34 & 12.52 & 2.53 & 11.02 & 4.37&21.2 ~5.6& 0.50 \\
       FCB            &  97 & -11.99 & 0.80 & -23.00 & 0.81 & -9.22 & 0.49 & 40.92 & 7.86 & 29.31 & 7.97 & 21.45 & 4.80&54.7 12.2& 5.47 \\
\hline
\end{tabular}
}  
\end{table*}

\subsection{Possible MG members among W UMa systems}

Moving groups (MGs) are kinematically coherent groups of stars that share a 
common origin. Eggen\ (1994) defined a supercluster of stars gravitationally 
unbound in the solar neighborhood, but sharing the same kinematics while 
occupying the extended regions in the Galaxy. Therefore, a MG, unlike the well 
known open clusters covering a limited region, can be observed all over the sky. 
The kinematical criteria, 
which were originally defined by Eggen\ (1958a, b, 1989, 1995), for determining 
the possible members of the best-documented MGs are summarized by Montes et al.\ 
(2001a, b). Evidence has been found that many young and active late-type binaries 
can be the members of some young moving groups (Jeffries\ 1995, Montes et al.\ 
2001b, King et al.\ 2003). Indeed, possible moving group members determined by 
Eggen's kinematical criteria have been proved to be very useful for separating 
kinematically heterogeneous CAB sample into two sub samples representing younger 
and older populations (Karata\c{s} et al.\ 2004) better than the classical 
approach of pre determined sub groups with a dividing line of a chosen parameter.

The difficulty of separating kinematically young and old populations in the 
velocity space is obvious. The dispersions increase with age but there are 
always some stars naturally occupying the regions near the LSR. It is, therefore, 
not safe to pick stars randomly near the LSR and then to form a kinematically 
young group with them. Eggen's kinematical criteria, however, objectively select 
some stars which have space velocity vector parallel to the converging point of 
each pre-determined and well known MGs within an acceptable uncertainty. There 
are two criteria that one sets a range of the direction of the test star's proper 
motion and other puts limits on the test stars radial velocity vector. 
Fulfilling one of the criteria makes the test star a possible member. But, 
fulfilling both criteria does not guarantee membership since there is 
always a possibility that the same velocity space could be occupied by the 
MG members and the non-members. Further independent criteria implying a common 
origin and same age as the member stars are needed to confirm the true membership. 
Therefore, we should always remember that Eggen's criteria determines only the 
possible members of a MG. The parameters of the five best-documented 
MGs and the possible membership criteria to them have been summarized by 
Karata\c{s} et al.\ (2004) for the CAB binaries. Here we apply the same criteria to 
the stars in the present W UMa sample and 28 systems were found to satisfy 
at least one of the criteria for one of the five MGs which were listed in 
Table 4. The possible moving group members among the W UMa binaries in the 
present sample is marked in the latest columns of Table 2 with the names of MG 
associated.

After determining the possible moving groups, the whole sample has been divided 
into three sub groups. The first one contains the possible moving group members 
and named MG. The second group contains only five stars which are previously 
known to be the members of well known open clusters. The third one is called 
'Field Contact Binaries' (FCB) which contains field W UMa systems free from the 
possible MG and the known open cluster members in the solar neighborhood. Fig. 7 
compares the $U-V$ diagram of the MG and the FCB. The detailed kinematics and 
implied kinematical ages of MG and FCB are included in Table 3.   

\begin{figure}
\resizebox{8.20cm}{14cm}{\includegraphics*{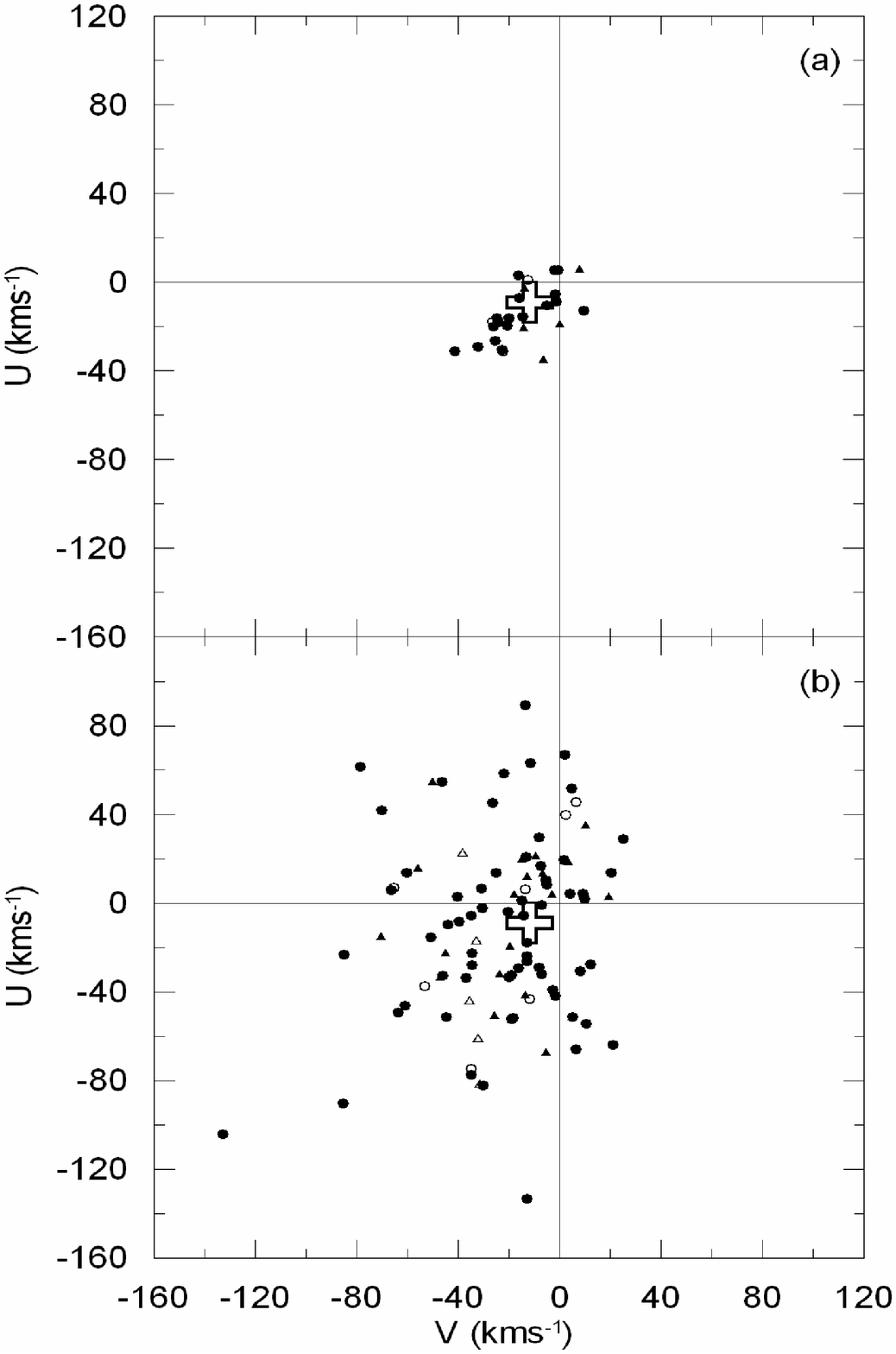}} 
\caption{Distribution of (a) possible MGs and (b) FCBs on the (U, V) 
diagram. The position of LSR is marked by +. Symbols are the same as Fig. 1.}
\end{figure}

The velocity dispersions of field contact binaries (FCB) imply a mean 
kinematical age of 5.47 Gyr. The mean kinematical age is found to be 500 Myr for 
the possible MG members. Although, this age appears to be consistent with the 
ages of MG in Table 4, it may not represent the true mean kinematical ages of all 
possible MG members. This is because, the ages in Table 3 are estimated from 
the dispersions with respect to the LSR. In reality, each MG has its own center 
of dispersion which is slightly different than the LSR. Therefore, we should 
keep in mind that members should be at the age of each MG listed in Table 4. 

\setcounter{table}{3}
\begin{table*}
\caption{The parameters of the best documented moving groups.} 
\begin{tabular}{lcccc}
\hline
Name &  Age &  ($U$, $V$, $W$) &    $V_{T}~^{*}$ &  C. P.$^{**}$ \\
     &(Myr) &     (km s$^{-1}$) & (km s$^{-1}$) & ($\alpha^{h}$, $\delta^{o}$)  \\
\hline
Local Association     & 20 -- 150 & (-11.6,-21.0,-11.4)& 26.5 & (5.98,-35.15) \\
(Pleiades, $\alpha$ Per, M34,&    &                    &      &               \\
$\delta$ Lyr, NGC 2516, IC 2602)   &                    &      &               \\
IC 2391 Supercluster  &  35 -- 55 & (-20.6,-15.7,-9.1) & 27.4 & (5.82,-12.44) \\
 (IC 2391)            &           &                    &      &               \\
 Castor MG            &    200    & (-10.7,-8.0,-9.7)  & 16.5 & (4.75,-18.44) \\
Ursa Major Group      &    300    & (14.9,1.0,-10.7)   & 18.4 & (20.55,-38.10)\\
(Sirius Supercluster) &           &                    &      &               \\
Hyades Supercluster   &    600    & (-39.7,-17.7,-2.4) & 43.5 & (6.40,6.50)   \\
(Hyades, Praesepe)    &           &                    &      &               \\
\hline
\end{tabular}
\\
\leftline{$^{*}$  Space velocity vector of MG as a whole ($V_{T}=\sqrt{U^{2}+V^{2}+W^{2}}$).}
\leftline{$^{**}$ Direction of space velocity vector (C. P.: Converging Point in celestial coordinates).}  
\end{table*}

\subsection{Comparing physical parameters between FCB and MG}

The kinematical criteria select out kinematically young W UMa systems 
without a dividing line according to any physical parameter of W UMa 
characteristics. However, this does not mean all kinematically young W UMa's 
are removed from the main sample when forming the group MG. There could be 
systems as young as the MG group in the FCB. Those are the systems which fail 
to satisfy kinematical criteria although they are young. Therefore young and 
old all sorts of W UMa system may exits in FCB. Thus, the 5.47 Gyr kinematical 
age represents an average age of all stars in FCB. On the other hand, 
One must remember that the MG group stars are not young because of their 
small dispersions on the velocity space. On the contrary, their dispersions 
are small because they are young. They are young primarily because they are 
possible members of young moving groups in Table 4.  Nevertheless, the 
word "possible" implies a limited number of non-members may well exist in 
MG which we believe not changes the statistic of young W UMa phenomenon.

Comparing the MG group and field CAB's, Karata\c{s} et al.\ (2004) found some 
important clues indicating mass loss and orbital period decrease in the binary 
evolution. A similar study is intended here to study the contact stage of the 
binary evolution. The physical parameters of W UMa binaries are listed in Table 5. 
The columns are self explanatory to indicate order number, name, 
apparent brightness and color ($B-V$), spectral type, binary type (A or W), 
orbital period and inclination, masses of components, mass ratio, and 
over-contact parameter $f$. The data was primarily collected from the same literature 
where the radial velocities were taken. Moreover, the orbital inclinations and $f$ 
values are taken from Pribulla, Kreiner, \& Tremko\ (2003) and Selam\ (2004).

The distribution according to spectral types are compared between the MG 
(younger) group and FCB (older) group in Fig. 8. It is interesting that 
both groups have a similar shape distribution. A slightly increased 
percentage of A type stars in MG group appears to be the only difference. 
Fig. 9 compares the orbital period distributions between MG and FCB. If one 
of the stars with an orbital period greater than 1 day in MG group is not 
counted, both groups cover nearly the same orbital period range ($0.2<P<0.9$ 
days). Both groups can be said to have a peak at $P=0.4$ days. With 27 members 
the MG group has an average period $<P>=0.5258$ days. With 97 members FCB group 
has an average period $<P>=0.4275$ days. Comparing the shape of both 
histograms we can conclude that the relative number of longer period ($P>0.4$ 
days) systems decreased and the relative number of shorter period systems 
($P<0.4$ days) increased in the older population sample (FCB). 

\begin{table*}
\caption{Physical parameters of W UMas.}
{\scriptsize
\begin{tabular}{crccccccccccc}
\hline
ID & Name &    $V$ &   $B-V$ & Spectral & Binary &    $P$  &   $i$      & $M_{1}$     & $M_{2}$ &  $q$ &      $f$ & Third/Multi\\
   &      &        &         &  Type    &  Type  &   (days)&  $(^{o})$  & $(M_{\odot})$ & $(M_{\odot})$&    &   & Component   \\

\hline
  1 &     AQ Tuc &       9.91 &       0.35 &       F2/5 &         A  &     0.5948 &       80.2 &      1.930 &      0.690 &      0.358 & 0.44 &\\
  2 &     BH Cas &      12.30 &       1.02 &        K4V &         W  &     0.4059 &       70.1 &      0.743 &      0.352 &      0.474 & 0.09 &\\
  3 &     DZ Psc &      10.18 &       0.51 &        F7V &         A  &     0.3661 &            &            &            &      0.136 &      &\\
  4 &   V523 Cas &      10.62 &       1.05 &        K4V &         W  &     0.2337 &       82.0 &      0.637 &      0.319 &      0.501 & 0.20&$\surd$\\
  5 &     AQ Psc &       8.68 &       0.54 &        F8V &         A  &     0.4756 &            &            &            &      0.226 &      &\\
  6 &     AE Phe &       7.56 &       0.64 &        F8V &         W  &     0.3624 &       85.3 &      1.366 &      0.629 &      0.461 & 0.17 &\\
  7 &     TW Cet &      10.43 &       0.67 &        G5V &         W  &     0.3117 &       83.2 &      1.284 &      0.680 &      0.530 & 0.03 &\\
  8 &   V776 Cas &       8.94 &       0.52 &        F2V &         A  &     0.4404 &       52.5 &      1.723 &      0.225 &      0.130 & 0.70 &\\
  9 &     QX And &      11.25 &       0.44 &        F5V &         A  &     0.4118 &       58.6 &      1.176 &      0.236 &      0.261 & 0.22 &\\
 10 &     SS Ari &      10.37 &       0.62 &        G0V &         W  &     0.4060 &       75.3 &      1.343 &      0.406 &      0.302 & 0.13&$\surd$\\
 11 &     GZ And &      10.83 &       0.78 &        G5V &         W  &     0.3050 &       85.9 &      1.170 &      0.602 &      0.514 & 0.03 &\\
 12 &   V376 And &       7.68 &       0.26 &        A4V &         A  &     0.7987 &            &            &            &      0.305 &      &\\
 13 &     EE Cet &       8.76 &       0.36 &        F2V &         W  &     0.3799 &       77.5 &      1.391 &      0.438 &      0.315 & 0.20 &\\
 14 &     UX Eri &      10.50 &       0.67 &        F9V &         A  &     0.4453 &       79.0 &      1.366 &      0.510 &      0.373 & 0.00 &\\
 15 &     EQ Tau &      10.50 &       0.83 &        G2V &         A  &     0.3413 &       86.6 &      1.217 &      0.537 &      0.442 & 0.12&$\surd$\\
 16 &     BL Eri &      11.50 &       0.63 &        G0V &         W  &     0.4169 &       88.0 &      0.609 &      0.330 &      0.542 & 0.16 &\\
 17 &     YY Eri &       8.10 &       0.67 &        G5V &         W  &     0.3215 &       82.5 &      1.540 &      0.620 &      0.403 & 0.12&$\surd$\\
 18 &     AO Cam &       9.50 &       0.68 &        G0V &         W  &     0.3299 &       75.1 &      1.189 &      0.491 &      0.413 & 0.10 &\\
 19 &     RZ Tau &      10.08 &       0.58 &        A7V &         A  &     0.4157 &       82.7 &      1.634 &      0.882 &      0.540 & 0.33 &\\
 20 &     DN Cam &       8.23 &       0.34 &        F2V &         W  &     0.4983 &       71.9 &      1.915 &      0.805 &      0.421 & 0.50 &\\
 21 &   V410 Aur &      10.10 &       0.64 &      G0/2V &         A  &     0.3663 &       90.0 &      1.234 &      0.178 &      0.144 & 0.50 &\\
 22 &   V402 Aur &       8.84 &       0.40 &        F2V &         W  &     0.6035 &            &            &            &      0.201 &      &\\
 23 &     AP Dor &       9.38 &       0.45 &        F5V &         W  &     0.4272 &       62.5 &            &            &      0.100 & 1.00 &\\
 24 &  V1363 Ori &      10.30 &       0.58 &        F8V &         A  &     0.4319 &            &            &            &      0.205 & 0.40 &\\
 25 &     ER Ori &       9.28 &       0.49 &        F7V &         W  &     0.4234 &       87.5 &      1.385 &      0.765 &      0.552 & 0.15&$\surd$\\
 26 &     RW Dor &      10.80 &       0.89 &        K1V &         W  &     0.2854 &       76.6 &      0.640 &      0.430 &      0.672 & 0.10 &\\
 27 &   V781 Tau &       8.90 &       0.57 &        G0V &         W  &     0.3449 &       65.4 &      1.237 &      0.501 &      0.405 & 0.23 &\\
 28 &     AH Aur &      10.20 &       0.55 &        F7V &         A  &     0.4943 &       75.5 &      1.683 &      0.284 &      0.169 & 0.23 &\\
 29 &     QW Gem &      10.33 &       0.48 &        F8V &         W  &     0.3581 &       85.0 &      1.275 &      0.425 &      0.334 & 0.50 &\\
 30 &   V753 Mon &       8.30 &       0.36 &        A8V &         W  &     0.6770 &       74.0 &      1.673 &      1.623 &      0.970 & 0.00 &\\
 31 &     TY Pup &       8.40 &       0.41 &        F3V &         A  &     0.8192 &       81.6 &      2.200 &      0.720 &      0.327 & 0.63 &\\
 32 &     FG Hya &       9.90 &       0.64 &        G2V &         A  &     0.3278 &       86.8 &      1.414 &      0.157 &      0.112 & 0.74 &\\
 33 &     TX Cnc &      10.00 &       0.42 &        F8V &         W  &     0.3829 &       62.4 &      0.792 &      0.420 &      0.535 & 0.08 &\\
 34 &     AH Cnc &      13.31 &       0.53 &        F5V &         W  &     0.3604 &       68.0 &      1.290 &      0.540 &      0.419 & 0.41 &\\
 35 &     UV Lyn &       9.41 &       0.64 &        F6V &         W  &     0.4150 &       66.8 &      1.349 &      0.495 &      0.367 & 0.45 &\\
 36 &     FN Cam &       8.58 &       0.38 &        A9V &         A  &     0.6771 &       71.2 &      2.402 &      0.532 &      0.222 & 0.56 &\\
 37 &     EZ Hya &      10.48 &       0.82 &        G2V &         W  &     0.4498 &            &            &            &      0.252 &      &\\
 38 &      S Ant &       6.40 &       0.36 &        A9V &         A  &     0.6483 &       68.8 &      1.940 &      0.760 &      0.392 & 0.07 &\\
 39 &      W UMa &       7.75 &       0.62 &        F8V &         W  &     0.3336 &       86.0 &      1.190 &      0.570 &      0.479 & 0.32 &\\
 40 &     AA UMa &      10.88 &       0.60 &        F9V &         W  &     0.4680 &       80.3 &      1.419 &      0.773 &      0.545 & 0.15 &\\
 41 &     RT Lmi &      11.40 &       0.60 &        F7V &         A  &     0.3749 &       84.0 &      1.298 &      0.476 &      0.367 & 0.23 &\\
 42 &     XY Leo &       9.45 &       1.04 &        K0V &         W  &     0.2841 &       65.8 &      0.870 &      0.435 &      0.500 & 0.06&$\surd$\\
 43 &     XZ Leo &      10.60 &       0.38 &        A8V &         A  &     0.4877 &       73.3 &      1.892 &      0.659 &      0.348 & 0.02 &\\
 44 &      Y Sex &       9.02 &       0.48 &         F8 &         A  &     0.4198 &       76.8 &      1.210 &      0.220 &      0.182 & 0.00 &\\
 45 &     ET Leo &       9.55 &       0.62 &        G8V &         W  &     0.3465 &       47.5 &      0.835 &      0.285 &      0.342 & 0.10 &\\
 46 &     UZ Leo &       9.58 &       0.39 &        A9V &         A  &     0.6180 &       79.7 &      2.074 &      0.629 &      0.303 & 0.85 &\\
 47 &     EX Leo &       8.27 &       0.53 &        F6V &         A  &     0.4086 &       61.1 &      1.557 &      0.309 &      0.199 & 0.31 &\\
 48 &     VY Sex &       9.02 &       0.55 &       F9.5 &         W  &     0.4434 &            &            &            &      0.313 &      &\\
 49 &     AM Leo &       9.25 &       0.56 &        F5V &         W  &     0.3658 &       87.0 &      1.386 &      0.623 &      0.449 & 0.15&$\surd$\\
 50 &     VW Lmi &       8.03 &       0.41 &        F3V &         A  &     0.4775 &       72.5 &            &            &      0.250 & 0.40 &\\
 51 &     AP Leo &       9.32 &       0.50 &        F8V &         A  &     0.4304 &       77.0 &      1.477 &      0.439 &      0.297 & 0.23 &\\
 52 &     HN UMa &       9.90 &       0.59 &        F8V &         A  &     0.3826 &            &            &            &      0.140 & 0.20 &\\
 53 &     AW UMa &       6.84 &       0.37 &        F2V &         A  &     0.4387 &       79.8 &      1.280 &      0.090 &      0.070 & 0.85&$\surd$\\
 54 &     TV Mus &      11.00 &       0.82 &        F8V &         A  &     0.4457 &       78.9 &      1.316 &      0.157 &      0.119 & 0.87 &\\
 55 &   V752 Cen &       9.17 &       0.58 &        F7V &         W  &     0.3702 &       81.7 &      1.320 &      0.410 &      0.311 & 0.06 &\\
 56 &     AG Vir &       8.35 &       0.29 &        A8V &         A  &     0.6427 &       80.7 &      1.610 &      0.510 &      0.317 &      &\\
 57 &     HX UMa &       8.89 &       0.49 &        F4V &         A  &     0.3792 &       50.0 &      1.333 &      0.387 &      0.291 & 0.70 &\\
 58 &     CC Com &      11.30 &       1.24 &        K5V &         W  &     0.2211 &       90.0 &      0.690 &      0.360 &      0.522 & 0.20 &\\
 59 &     AH Vir &       8.89 &       0.81 &        G8V &         W  &     0.4075 &       86.5 &      1.360 &      0.412 &      0.303 & 0.24&$\surd$\\
 60 &     II UMa &       8.17 &       0.61 &      F5III &         A  &     0.8250 &       70.0 &      2.238 &      0.385 &      0.172 & 0.70 &\\
 61 &     RW Com &      11.07 &       0.84 &        K0V &         W  &     0.2373 &       75.2 &      0.920 &      0.310 &      0.337 & 0.17 &\\
 62 &     RZ Com &      10.42 &       0.52 &        F8V &         W  &     0.3385 &       86.0 &      1.108 &      0.484 &      0.437 & 0.07 &\\
 63 &     SX Crv &       9.07 &       0.56 &        F6V &         A  &     0.3166 &            &            &            &      0.066 &      &\\
 64 &     BI CVn &      10.26 &       0.62 &        F2V &         A  &     0.3842 &       69.2 &      1.646 &      0.679 &      0.413 & 0.17 &\\
 65 &     KZ Vir &       8.45 &       0.47 &        F7V &         A  &     1.1318 &            &            &            &      0.848 &      &\\
\hline
\end{tabular}  
}
\end{table*}

\begin{table*}
\contcaption{}
{\scriptsize
\begin{tabular}{crccccccccccc}
\hline
ID & Name &    $V$ &   $B-V$ & Spectral & Binary &    $P$  &   $i$      & $M_{1}$     & $M_{2}$ &  $q$ &      $f$ & Third/Multi\\
   &      &        &         &  Type    &  Type  &   (days)&  $(^{o})$  & $(M_{\odot})$ & $(M_{\odot})$&    &   & Component   \\
\hline
 66 &     KR Com &       7.26 &       0.57 &       G0IV &         A  &     0.4080 &            &            &            &      0.091 & 0.70 &\\
 67 &     HT Vir &       7.06 &       0.56 &        F8V &         A  &     0.4077 &       90.0 &      1.258 &      1.022 &      0.812 &      &\\
 68 &     XY Boo &      10.30 &       0.49 &        F5V &         A  &     0.3706 &       69.0 &      0.934 &      0.147 &      0.158 & 0.05 &\\
 69 &   V757 Cen &       8.30 &       0.64 &        F9V &         W  &     0.3432 &       69.3 &      1.000 &      0.690 &      0.690 & 0.14 &\\
 70 &     RR Cen &       7.27 &       0.34 &        A9V &         A  &     0.6057 &       78.7 &      1.854 &      0.389 &      0.210 & 0.35 &\\
 71 &     VW Boo &      10.50 &       0.81 &        G5V &         W  &     0.3422 &       76.2 &      0.980 &      0.420 &      0.428 & 0.72 &\\
 72 &     NN Vir &       7.64 &       0.41 &     F0/F1V &         A  &     0.4807 &       65.0 &      1.794 &      0.882 &      0.491 & 0.50 &\\
 73 &     EF Boo &       9.43 &       0.50 &        F5V &         W  &     0.4295 &       75.2 &      1.554 &      0.795 &      0.512 & 0.20 &\\
 74 &     CK Boo &       8.99 &       0.54 &      F7/8V &         A  &     0.3552 &       56.6 &      1.788 &      0.198 &      0.111 & 0.14 &\\
 75 &     GR Vir &       7.80 &       0.58 &      F7/8V &         A  &     0.3470 &            &            &            &      0.122 &      &\\
 76 &     AC Boo &      10.00 &       0.59 &       F8Vn &         W  &     0.3524 &       82.6 &      1.403 &      0.565 &      0.403 & 0.91 &\\
 77 &     TY Boo &      10.81 &       0.76 &        G5V &         W  &     0.3171 &       76.6 &      0.930 &      0.400 &      0.437 & 0.10&$\surd$\\
 78 &     44 Boo &       5.80 &       0.85 &        K2V &         W  &     0.2678 &       72.8 &      0.900 &      0.439 &      0.487 & 0.00&$\surd$\\
 79 &     TZ Boo &      10.41 &       0.63 &        G2V &         A  &     0.2972 &       66.9 &      0.783 &      0.104 &      0.133 &      &\\
 80 &     BW Dra &       8.61 &       0.63 &        F8V &         W  &     0.2923 &       74.4 &      0.891 &      0.250 &      0.281 & 0.14 &\\
 81 &     BV Dra &       7.88 &       0.56 &        F7V &         W  &     0.3501 &       76.3 &      0.997 &      0.401 &      0.402 & 0.11 &\\
 82 &     FI Boo &       9.60 &       0.72 &        G3V &         W  &     0.3900 &            &            &            &      0.372 & 0.10 &\\
 83 &     OU Ser &       8.25 &       0.64 &     F9/G0V &         A  &     0.2968 &       74.3 &      1.018 &      0.176 &      0.173 & 0.31 &\\
 84 &     VZ Lib &      10.13 &       0.64 &         G0 &         A  &     0.3583 &            &            &            &      0.237 &      &\\
 85 &     FU Dra &      10.55 &       0.59 &        F8V &         W  &     0.3067 &       78.6 &      1.168 &      0.293 &      0.251 & 0.23 &\\
 86 &     YY CrB &       8.64 &       0.62 &        F8V &         A  &     0.3766 &       77.0 &      1.429 &      0.348 &      0.243 & 0.63 &\\
 87 &     AU Ser &      10.90 &       0.87 &        K0V &         A  &     0.3865 &       80.6 &      0.921 &      0.646 &      0.701 & 0.09 &\\
 88 &   V842 Her &       9.85 &       0.65 &        F9V &         W  &     0.4190 &       79.0 &      1.360 &      0.353 &      0.260 & 0.25 &\\
 89 &     BX Dra &      10.50 &       0.35 &     F0IV-V &         A  &     0.5790 &            &            &            &      0.289 &      &\\
 90 &   V899 Her &       7.93 &       0.45 &        F5V &         A  &     0.4212 &       68.7 &      1.836 &      1.039 &      0.566 & 0.30 &\\
 91 &   V502 Oph &       8.34 &       0.65 &        G0V &         W  &     0.4534 &       71.3 &      1.297 &      0.481 &      0.335 & 0.24&$\surd$\\
 92 &   V918 Her &       7.30 &       0.28 &        A7V &         W  &     0.5748 &            &            &            &      0.271 &      &\\
 93 &   V921 Her &       9.44 &       0.39 &       A7IV &         A  &     0.8774 &       87.5 &      1.609 &      0.365 &      0.226 & 0.20 &\\
 94 &   V829 Her &      10.10 &       0.65 &        G2V &         W  &     0.3582 &       57.0 &      1.275 &      0.520 &      0.408 &      &\\
 95 &  V2357 Oph &      10.67 &       0.86 &        G5V &         A  &     0.4156 &            &            &            &      0.231 & 0.40 &\\
 96 &     AK Her &       8.29 &       0.55 &        F5V &         A  &     0.4215 &       80.8 &      1.310 &      0.300 &      0.229 & 0.10&$\surd$\\
 97 &   V728 Her &      10.90 &       0.41 &        F3V &         W  &     0.4713 &       69.2 &      1.654 &      0.295 &      0.178 & 0.71 &\\
 98 &     GM Dra &       8.77 &       0.51 &        F5V &         W  &     0.3387 &            &            &            &      0.180 & 0.40 &\\
 99 &  V2377 Oph &       8.60 &       0.66 &      G0/1V &         W  &     0.4254 &            &            &            &      0.395 & 0.80 &\\
100 &   V535 Ara &       7.17 &       0.34 &        A8V &         A  &     0.6293 &       82.1 &      1.520 &      0.460 &      0.303 & 0.03 &\\
101 &  V2388 Oph &       6.25 &       0.47 &        F3V &         A  &     0.8023 &       83.9 &      1.648 &      0.306 &      0.186 & 0.71 &\\
102 &   V566 Oph &       7.46 &       0.45 &        F4V &         A  &     0.4096 &       79.8 &      1.469 &      0.357 &      0.243 & 0.41 &\\
103 &   V972 Her &       6.72 &       0.47 &        F4V &         W  &     0.4431 &            &            &            &      0.167 & 0.00 &\\
104 &   V508 Oph &      10.06 &       0.67 &        F9V &         A  &     0.3448 &       86.1 &      1.010 &      0.520 &      0.515 & 0.10 &\\
105 &     EF Dra &      10.48 &       0.82 &        F9V &         A  &     0.4240 &       78.1 &      1.813 &      0.289 &      0.160 & 0.45&$\surd$\\
106 &   V839 Oph &       8.80 &       0.62 &        F7V &         A  &     0.4090 &       80.1 &      1.640 &      0.496 &      0.305 & 0.39 &\\
107 &    eps Cra &       4.74 &       0.40 &        F2V &         A  &     0.5914 &       72.3 &      1.720 &      0.220 &      0.128 & 0.30 &\\
108 &   V401 Cyg &      10.64 &       0.38 &        F0V &         A  &     0.5827 &       77.0 &      1.679 &      0.487 &      0.290 & 0.46 &\\
109 &  V2082 Cyg &       6.68 &       0.35 &        F2V &         A  &     0.7141 &            &            &            &      0.238 &      &\\
110 &   V417 Aql &      11.00 &       0.53 &        G2V &         W  &     0.3703 &       84.5 &      1.395 &      0.505 &      0.362 & 0.19 &\\
111 &     OO Aql &       9.20 &       0.61 &        G2V &         A  &     0.5068 &       90.0 &      1.040 &      0.880 &      0.846 & 0.27 &\\
112 &     VW Cep &       7.23 &       0.86 &        G9V &         W  &     0.2783 &       63.6 &      0.897 &      0.247 &      0.275 & 0.22&$\surd$\\
113 &     LS Del &       8.61 &       0.68 &        G0V &         W  &     0.3638 &       48.5 &      1.068 &      0.399 &      0.375 & 0.06 &\\
114 &  V2150 Cyg &       8.09 &       0.40 &        A6V &         A  &     0.5919 &            &            &            &      0.802 & 0.90 &\\
115 &     HV Aqr &       9.71 &       0.63 &        F5V &         A  &     0.3745 &       78.3 &      1.366 &      0.198 &      0.145 & 0.47 &\\
116 &  V1073 Cyg &       8.23 &       0.45 &        F2V &         A  &     0.7859 &       68.4 &      1.600 &      0.510 &      0.319 & 0.04 &\\
117 &     KP Peg &       7.07 &       0.16 &        A2V &         A  &     0.7272 &            &            &            &      0.322 &      &\\
118 &     DK Cyg &      10.37 &       0.39 &        A8V &         A  &     0.4707 &       80.3 &      1.768 &      0.575 &      0.325 & 0.55 &\\
119 &     BX Peg &      11.00 &       0.78 &       G4.5 &         W  &     0.2804 &       87.5 &      1.020 &      0.380 &      0.373 & 0.18 &\\
120 &   V445 Cep &       6.85 &       0.14 &        A2V &         A  &     0.4488 &            &            &            &      0.167 &      &\\
121 &     BB Peg &      10.80 &       0.52 &        F5V &         W  &     0.3615 &       79.9 &      1.431 &      0.515 &      0.360 & 0.37 &\\
122 &   V335 Peg &       7.24 &       0.50 &        F5V &         A  &     0.8107 &            &            &            &      0.262 & 0.10 &\\
123 &     SW Lac &       8.51 &       0.75 &     K0Vvar &         W  &     0.3207 &       80.2 &      0.959 &      0.779 &      0.851 & 0.39&$\surd$\\
124 &     AB And &       9.50 &       0.87 &        G8V &         W  &     0.3319 &       86.8 &      1.005 &      0.494 &      0.560 & 0.15&$\surd$\\
125 &   V351 Peg &       8.00 &       0.34 &        A8V &         W  &     0.5933 &       62.5 &      2.327 &      0.838 &      0.360 & 0.40 &\\
126 &     VZ Psc &      10.20 &       1.14 &        K5V &         A  &     0.2613 &       49.6 &      0.790 &      0.720 &      0.911 & 0.00 &\\
127 & BD +14 5016&       9.50 &       0.31 &        F0V &         A  &     0.6369 &       72.6 &      1.531 &      0.387 &      0.253 & 0.54 &\\
128 &     EL Aqr &      10.49 &       0.50 &        F3V &         A  &     0.4814 &       70.8 &      1.563 &      0.317 &      0.203 &      &\\
129 &      U Peg &       9.23 &       0.62 &        G2V &         W  &     0.3748 &       77.5 &      1.149 &      0.379 &      0.330 & 0.09 &\\
\hline&
\end{tabular}  
}
\end{table*}

\begin{figure}
\resizebox{8.20cm}{9.9cm}{\includegraphics*{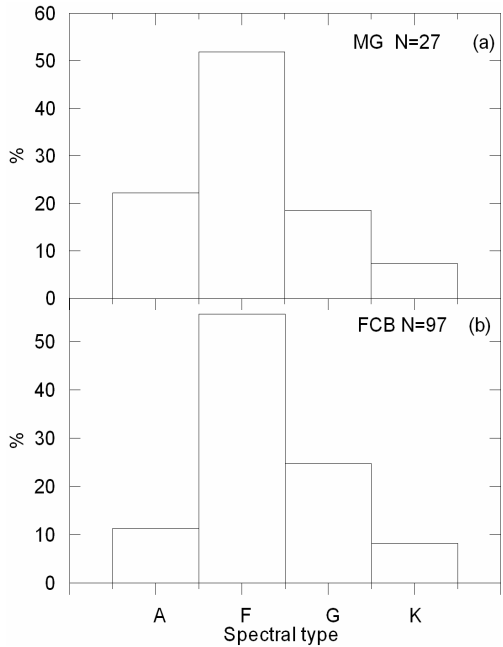}} 
\caption{Comparison of spectral type histograms of (a) MGs and (b) FCBs.}
\end{figure}

\begin{figure}
\resizebox{8.20cm}{9.9cm}{\includegraphics*{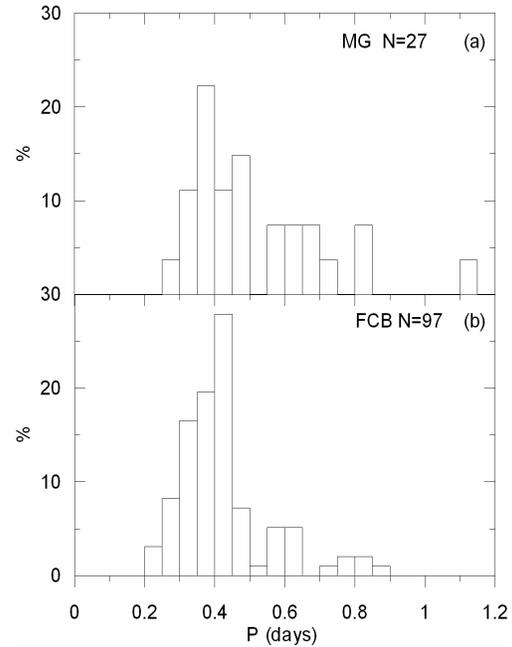}} 
\caption{Comparison of period histograms of (a) MGs and (b) FCBs.}
\end{figure}

These changes may appear as evidences of the period decrease in the contact 
binary evolution. Karata\c{s} et al.\ (2004) assumed the period histogram of 
possible MG members as the initial period distribution of field CAB. But, 
here, the situation is totally different. If the contact stage time scale is 
indeed short as estimated to be within the range $0.1<t_{contact}<1$ Gyr by 
Guinan \& Bradstreet\ (1988), MG contact binaries cannot be taken as an 
initial distribution of the FCB, because MG contact binaries will not be 
able to survive up to the mean kinematical age of FCB. 

The mean kinematical age of FCB is 5.47 Gyr (Table 3). Karata\c{s} 
et al.\ (2004) assigned 3.86 Gyr age for the field CAB which are 
potential progenitors of field contact binaries (Demircan\ 1999). Therefore, 
if there is a continuous formation of CAB and evolution out of the CAB stage,
some (or maybe most) of them become W UMa systems, and similarly, if the formation 
of W UMa systems and their evolution (coalescence) into a single star is a continuous 
process, and somehow if there is an equilibrium in the Galaxy, which could be 
investigated of course, the 1.61 Gyr age difference between field CAB and field FCB 
implies an upper limit for the timescale of the contact stage. Indeed, it is not much 
different than the estimate of $0.1<t_{contact}<1$ Gyr by Guinan \& Bradstreet\ (1988).

Therefore, it is not possible to assume the MG group to be the initial stage of the 
FCB group. It is most likely that the MG contact systems are formed in the beginning 
of the main-sequence or during the pre-main-sequence contraction phase, either by a 
fission process (Roxburgh\ 1966) or most probably by fast spiraling in of two 
components in a common envelope. On the other hand, FCBs must be mostly formed 
from the detached progenitors.

The evidence of total mass and orbital period decrease appears to be better 
displayed in Table 6 where the field contact binaries are divided into four 
sub groups according to orbital period criterion. It is clearly displayed there that 
the average kinematical age increases as the orbital periods of the sub groups 
decrease. It is also clear that older sub groups have smaller total mass.
As if, long period contact binaries in Table 6 lose mass and angular momentum, so 
their orbital period decrease within the time scale indicated in the Table. 
This scenario, however, is definitely misleading and cannot be true. This is 
because the age difference between the youngest and oldest sub groups in Table 6, 
is too long for a contact system to survive even if the empirical 1.61 Gyr 
time scale for the contact stage were taken into consideration.     

According to the true scenario, the large dispersions (older ages) in Table 6 must 
have been produced not only during the contact stage, but also 
including the pre-contact stages before, which may be longer than the contact stage 
as indicated by Table 6. Thus, one should never expect, for a W UMa system in the 
youngest group in Table 6, to loose mass and angular momentum and join the shortest 
period (oldest) W UMa systems after $t=8.89-3.21=5.68$ Gyr later. It is not yet clear, 
but it seems that when a detached system becomes a contact binary, the angular momentum 
loss becomes accelerated with the gravitational radiation as described by Guinan \& 
Bradstreet\ (1988). If the total mass of the system does not decrease to fit in the 
decreasing Roche lobes, then, overcontact $f$ parameter starts to increase. The 
pre-contact stage, which may differ one system to system, being much longer than the 
contact stage, spoils the correlation between the $f$ parameter and the space velocity 
as described in Fig. 5.  

\begin{figure}
\resizebox{8.20cm}{9.9cm}{\includegraphics*{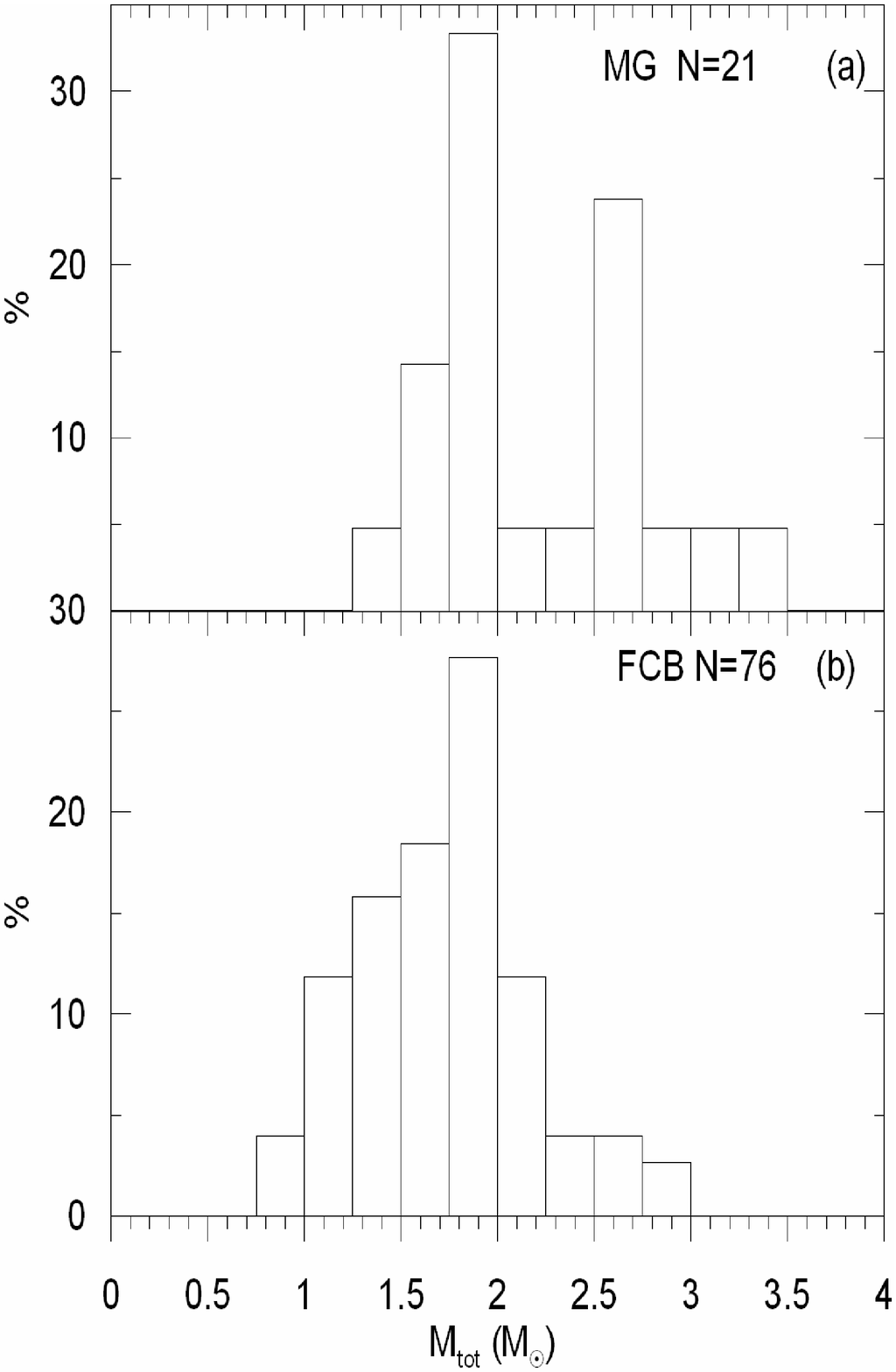}} 
\caption{Comparison of the total mass ($M_{h}$ + $M_{c}$) histograms of (a) 
MGs and (b) FCBs.}
\end{figure}
   
\setcounter{table}{5}
\begin{table}
\caption{Kinematical ages for the period sub-groups in FCB.} 
\begin{tabular}{cccccc}
\hline
      $P$  &$<M_{tot}>$ &   $<P>$  &   N &  $\sigma_{tot}$ & Age \\
    (days) &($M_\odot$) &  (days)    &     &(km s$^{-1}$) & (Gyr) \\
\hline
(0.5-0.9] &      2.220 &      0.669 &  17 &       42.09~~~7.1 & 3.21 \\
(0.4-0.5] &      1.900 &      0.434 &  34 &       43.86~~10.4 & 3.51 \\
(0.3-0.4] &      1.618 &      0.353 &  35 &       63.90~~16.0 & 7.14 \\ 
(0.2-0.3] &      1.171 &      0.270 &  11 &       73.92~~14.6 & 8.89 \\
\hline
\end{tabular}  
\end{table}

Fig. 10 displays the total mass histograms of possible MG and FCB groups. 
Similar to the period histograms that it would be meaningless if one tries to 
deduce evidence of mass decrease during contact binary evolution from this 
figure. Therefore, it should be considered as a bare comparison between MG and FCB. It 
is the same with Fig. 11, which compares the mass ratio distribution between MG 
and the FCB groups. The mass ratio evolution during the contact stage cannot 
be deduced from Fig. 11 since MG group cannot be the initial distribution. 
On the other hand, all four histograms (Figs. 8, 9, 10, and 11) display an empirical 
evidence of contact binary evolution to indicate coalescence of W UMa systems into 
single stars. Otherwise, angular momentum loss does not fit in the theory of binary 
evolution. Then, it becomes difficult to explain why there are such differences between 
the MG and the FCB group W UMa stars.   

\begin{figure}
\resizebox{8.20cm}{9.9cm}{\includegraphics*{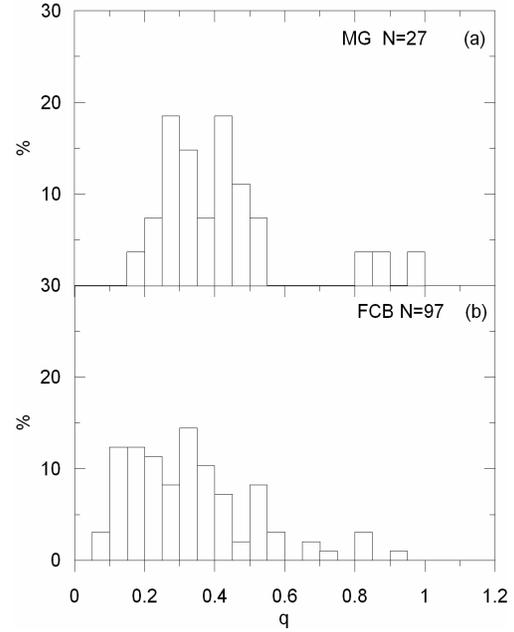}} 
\caption{Comparison of the mass ratio ($q=M_{2}/M_{1}$, where $M_{2} < M_{1}$, and 
 $M_{1}$, $M_{2}$ are the primary and secondary masses, respectively) histograms of 
(a) MGs and (b) FCBs.}
\end{figure}

A similar case also applies to the Table 6 data. We believe the age difference 
between MG contact binaries and the youngest FCB group in Table 6 is empirical 
evidence of coalescence into single stars. This is because the coalescence into a single star 
involves orbital period decrease as well as angular momentum loss, most probably 
accelerated by gravitational radiation as described by Guinan \& Bradstreet\ 
(1988) although the data in Table 6 is mostly affected by the long pre-contact stages. The 
derivation of the rates of orbital period decrease, mass and angular momentum 
loss from the present data of contact binaries requires a careful analysis and 
interpretation of Table 6 data. Due to limited space in this study derivation of 
those rates will be handled in a forthcoming study. 

\section{Conclusion}

Kinematics of 129 W UMa binaries is studied. The sample is found to be 
heterogeneous in the velocity space that kinematically younger and older 
contact binaries exist in the present sample. Various sub groups are formed 
according to criteria involving the spectral types, binary type (A or W), mass ratio, and 
over-contact parameter $f$, orbital period and total mass. However, those criteria failed 
to produce kinematically homogeneous sub groups. Nevertheless, a correlation has been 
found to indicate shorter period and less massive systems show larger velocity 
dispersions than longer period more massive systems. The mean kinematical ages of all 
sub groups are estimated from the velocity dispersions using the tables of Wielen\ (1982) 
and listed in Table 3.

The possible moving group members among the present W UMa stars are 
investigated according to the kinematical criteria originally defined by Eggen\ 
(1958a, b, 1989, 1995) and 28 W UMa are found to be a possible member of five 
best documented MGs which are summarized in Table 4. The group of the field contact 
binaries (FCB) is formed by removing out the possible moving group members and the 
W UMa systems which are known to be the members of known open clusters from the main 
sample. The 5.47 Gyr mean kinematical age is estimated for the field contact systems (FCB). 
The age difference is found to be 1.61 Gyr between the 3.86 Gyr old CAB (Karata\c{s} 
et al.\ 2004) and the FCB. If there is a number equilibrium between the CAB and FCB 
established by the evolution, the upper limit of the lifetime of the contact stage can 
be estimated as 1.61 Gyr which is in close agreement with the estimation 
made by Guinan \& Bradstreet\ (1988) as $0.1<t_{contact}<1$ Gyr.

Since the contact stage appears to be very short in comparison with the mean 
kinematical age of FCB, the group containing possible MG stars cannot be used 
as an initial sub sample of FCB. Thus, the 
histograms (Figs. 8, 9, and 10) and ages of sub groups of FCB according to 
orbital ranges (Table 6) cannot be used directly to deduce evidence of period 
decrease and angular momentum loss during contact stage evolution. Nevertheless,        
since, those histograms and data in Table 6, cannot be explained without angular 
momentum loss and orbital period decrease in binary evolution, and since it is 
illogical to assume that angular momentum loss and orbital period decrease operate 
only for the pre-contact stages, the angular momentum loss and the orbital period 
decrease are evidently occuring as predicted by Guinan \& Bradstreet\ (1988) even 
if one cannot deduce them directly from those histograms and Table 6. 

We have found 27 systems to form the MG group. Existences of MG members is an 
exciting (perhaps the most exciting) finding of this study. This is because these 
W UMa systems do not have enough ages to be formed from detached binary progenitors 
by the mechanism involving angular momentum loss and orbital period decrease. In 
fact the age difference between the youngest FCB sub group (3.21 Gyr, in Table 6) 
and MG group (0.5 Gyr, in Table 3) is more than lifetime (1.61 Gyr) of the contact stage. 
Therefore, it could be concluded that the MG contact systems are formed in the 
beginning of the main-sequence or during the pre-main-sequence contraction phase, 
either by a fission process (Roxburgh\ 1966) or by fast spiraling the two components 
in a common envelope. As a result, it can be said that detached binary progenitors, 
presumably short period ($P<5$ days) RS CVn binaries, are not the only source from 
which W UMa systems could be formed by the mechanism involving angular momentum loss 
and orbital period decrease. It is indeed a challenge to disprove true membership 
of MG stars in our list. This may not be enough, one also needs to prove their 
ages to be bigger than our estimate, that is, their ages must be proven to be big 
enough (bigger than the pre-main sequence contraction plus the contact stage 
life time) then one can claim the detached binary progenitors are the only 
source to form W UMa systems. 

\section{Acknowledgments}
This research has made use of the SIMBAD database, operated at CDS, Strasbourg, 
France. Thanks to B. J. Hrivnak for providing private communications.
We also thank to Dr. Slavek Rucinski for his substantial input into the paper as 
the referee. A British national, who is an English teacher, Margaret Kad\i o\u glu 
helped on the grammar and linguistics. This work was supported by the Research 
Fund of the University of Istanbul. Project number 244/23082004.

\end{document}